\documentclass[11pt,a4paper]{article}
\usepackage{jheppub}
\usepackage[english]{babel}
\usepackage{exscale,color}
\usepackage[utf8]{inputenc}
\usepackage{amsfonts}
\usepackage{amsthm}
\usepackage{mathrsfs}
\usepackage{xcolor}
\usepackage{caption}
\usepackage{subcaption}
\usepackage{dcolumn}
\usepackage{bm}
\usepackage{multirow}
\allowdisplaybreaks

\title{Enhancement of the double Higgs production via leptoquarks at  the LHC}

\author[a]{Leandro Da Rold,}
\author[b]{Manuel Epele,}
\author[c]{Anibal Medina,}
\author[c]{Nicol\'as I. Mileo}
\author[c]{and Alejandro Szynkman}

\affiliation[a]{Centro At\'omico Bariloche, Instituto Balseiro and CONICET,\\
Av. Bustillo 9500, 8400, S. C. de Bariloche, Argentina}
\affiliation[b]{University of Jyvaskyla, Department of Physics,\\
P.O. Box 35, FI-40014 University of Jyvaskyla, Finland}
\affiliation[c]{IFLP, CONICET - Dpto. de F\'{\i}sica, Universidad Nacional de La Plata,\\ 
C.C. 67, 1900 La Plata, Argentina}

\emailAdd{daroldl@ib.edu.ar}
\emailAdd{manuel.m.epele@jyu.fi}
\emailAdd{anibal.medina@fisica.unlp.edu.ar}
\emailAdd{mileo@fisica.unlp.edu.ar}
\emailAdd{szynkman@fisica.unlp.edu.ar}

\abstract{Measurements of single Higgs production and its decays are in good agreement with the Standard Model. There is still room for large modifications in double Higgs production at LHC, though these effects may be correlated with large corrections to other observables, in particular single Higgs production. In this work we address the issue of enhancing double Higgs production in the presence of scalar leptoquarks while satisfying all experimental constraints. We show at leading order that including more than one species of leptoquarks, large cubic interactions with the Higgs can lead to sizable enhancement of di-Higgs production cross section at LHC, while at the same time keeping other Higgs observables and precision measurements under control. For masses above 800 GeV these corrections are in general below 30\%, whereas in a viable scenario in which one of the leptoquarks can be light, specifically in the mass range $400-600$ GeV, we show that it is possible to roughly double the SM cross section for di-Higgs production, implying that possible first hints of it may be probed at the high luminosity LHC at $\mathcal{L}\sim 2$ ab$^{-1}$.}


\arxivnumber{2105.06309}

\begin{document}
\maketitle

\section{Introduction}

A particle which highly resembles the Higgs boson of the Standard Model (SM) of particle physics has been discovered at the LHC in 2012~\cite{Aad:2012tfa,Chatrchyan:2012ufa} and since then, many of its couplings to the rest of the SM spectrum have been constrained by studying its single production and dominant decays~\cite{Khachatryan:2016vau,Sopczak:2020vrs}, pointing towards its SM-like nature. In fact no sign of new physics has been discovered at the LHC by direct or indirect probes so far, but nonetheless there still remains room for the possibility that some of the Higgs interactions may be affected by new physics.

We consider in this work an example of the last situation, how the double Higgs production process at the LHC can possibly be enhanced in a model of beyond the SM physics (BSM) that contains scalar leptoquarks (LQ) which interact with the Higgs and SM fermions. Leptoquarks have been also considered in connection with the phenomenology of the Higgs boson in Refs.~\cite{DaRold:2018moy,Zhang:2019jwp,Bhaskar:2020kdr,Crivellin:2020ukd}. It is well known that within the SM and in the present experiments, the multiple Higgs production process is the only way to access the interactions provided by the potential responsible for the spontaneous breaking of the Electroweak (EW) gauge symmetry. However, double Higgs production cross section in the SM case will only be probed at the high luminosity LHC ($\mathcal{L}\sim 3000$~fb$^{-1}$)~\cite{deFlorian:2015moa,Grazzini:2018bsd,deFlorian:2019app,Cepeda:2019klc}. On the other hand, LQs can run in the loops of gluon fusion diagrams since they are scalars that carry color charge and may also posses large cubic and quartic interactions that involve the Higgs, allowing an enhancement in the double Higgs production~\cite{Enkhbat:2013oba}.

LQs have also recently been advocated as some of the best candidates to explain leptonic anomalies, as the anomalous magnetic moment of the muon~\cite{Bennett:2006fi,Abi:2021gix}~\footnote{Notice, however, that a recent lattice calculation alleviates the tension between the combined experimental result and the SM theoretical prediction~\cite{Borsanyi:2020mff}.} and decays of $B$-mesons~\cite{Lees:2013uzd,Hirose:2016wfn,Aaij:2015yra,Aaij:2014ora,Aaij:2017vbb,Aaij:2019wad,Abdesselam:2019wac,Aaij:2021vac}. However, due to the leptoquark nature, they are usually involved in many flavor changing transitions which constrain couplings, mostly to 1st and 2nd generations. There are also strong constraints for LQs coming from their direct QCD production and subsequent decay into SM fermions as well indirect constraints to some of the possible cubic and quartic interactions  allowed in the scalar sector due to $T$-parameter contributions and potential contributions to the $Zb\bar{b}$-coupling~\cite{Dorsner:2016wpm}. Since, as we will show in the following sections, one needs large values of the cubic interactions involving Higgs and LQs in order to enhance double Higgs production, there are perturbativity limits on how large these can be, as well as possible implications in a destabilization of the electroweak vacuum, both of which we address. Finally there can be a strong tension in enhancing double Higgs production and at the same time keeping under control the contributions to single Higgs production, a delicate problem on which a central part of this work is dedicated to.

We take into account the bounds described in the previous paragraph and, by performing a numerical study which in no way is intended to be thorough, we show that it is indeed possible in our BSM model to obtain cross sections for double Higgs production which  can be slightly larger than 2 times the SM one, while at the same time satisfying all experimental constraints. This provides for an opportunity of observing double Higgs production at LHC at a lower luminosity than the ones required for the SM case, possibly reaching discovery levels at $\mathcal{L}< 3000$~fb$^{-1}$.

The paper is organized as follows. In Sec.~\ref{model} we introduce the model of LQs that we will use in order to enhance the double Higgs production at the LHC, considering stability constraints, working out the physical states and their mixing, and writing out the LQs interactions with fermions. Sec.~\ref{higgs phenomenology} deals with the proper calculation of the double Higgs production at the LHC in our model, taking into account as well the contributions to single Higgs production which constrain the enhancement. In this section we also introduce a very useful limit known as the low energy theorem (LET), in which the interactions between the Higgs and gluons take a simple form that allow us to understand better the physical behavior in certain regions of the parameter space. In Sec.~\ref{constraints} we focus on the constraints on the model from possible oblique corrections to electroweak precision tests (EWPT), LQs contributions to the well measured value of the $Zb\bar{b}$ coupling, flavor changing transitions and LQ direct searches. We show and analyze our numerical results in Sec.~\ref{sec-light-scenario} in the case in which one of the LQ remains light, whereas Sec.~\ref{sec-heavy-scenario} is devoted to the heavy LQs scenario. In Sec.~\ref{Banomalies} we briefly comment on how some of our LQs could address the recently $B$-anomalies measurements. Finally we give our conclusions in Sec.~\ref{conclusions}.

\section{Model}\label{model}
The aim of our work consists in constructing and studying, from a phenomenological bottom-up approach, a model beyond the SM that includes only a small number of LQs which provide an enhancement in the double Higgs production at the LHC. In that spirit, we consider a theory in which we add to the SM content the following three scalar LQs following the notation used in~\cite{Dorsner:2016wpm}
\begin{equation}
\tilde R_2\sim({\bf 3},{\bf 2})_{1/6}\ ,\qquad
S_1\sim(\bar{\bf 3},{\bf 1})_{1/3}\ ,\qquad
\bar S_1\sim(\bar{\bf 3},{\bf 1})_{-2/3}\ ,
\end{equation}
where the numbers in parenthesis represent the $SU(3)_C \times SU(2)_L$ SM gauge group transformation properties, and the subindex the LQ hypercharge. We denote generically the scalar fields as: $\phi=H,\tilde R_2,S_1,\bar S_1$, and use greek: $\alpha, \beta, \dots$, and latin: $i,j,\dots$ indices for color and weak isospin in the fundamental representation, respectively. The reason behind the choice of these particular LQs becomes clearer once we write the most general renormalizable potential, and in particular the possible cubic and quartic interactions between the LQs and the SM Higgs~\cite{Dorsner:2016wpm,Hirsch:1996qy}
\begin{equation}
V=V_2+V_3+V_4 \ ,
\end{equation}
with
\begin{align}
V_2=&\sum_\phi m_\phi^2 \phi^2\ , \label{eq-V2}\\
V_3=&\mu_1 H^\dagger \tilde R_2 S_1 +\mu_2 \tilde R_2^j \epsilon_{jk} H^k\bar S_1 +{\rm h.c.} \ , \label{eq-V3}\\
V_4=&\sum_\phi \lambda_\phi (\phi^\dagger\phi)^2+\lambda_1 H^\dagger H\tilde R_2^\dagger\tilde R_2+\lambda_2 H^\dagger HS_1^\dagger S_1+\lambda_3 H^\dagger H\bar S_1^\dagger\bar S_1+\lambda_4 \bar S_1^\dagger \bar S_1\tilde R_2^\dagger\tilde R_2 \, + \nonumber\\
&\lambda_5 S_1^\dagger S_1\tilde R_2^\dagger\tilde R_2+\lambda_6 S_1^\dagger S_1\bar S_1^\dagger\bar S_1 + \lambda'_1 H^\dagger\sigma_i H\tilde R_2^\dagger\sigma_i\tilde R_2 + \lambda'_{\tilde R_2}\tilde R_2^\dagger\sigma_i t^a\tilde R_2\tilde R_2^\dagger\sigma_i t^a\tilde R_2 \, + \nonumber\\
&\lambda_7\tilde R_2^{\alpha i} \tilde R_2^{\beta j} S_1^\delta\bar S_1^\rho\epsilon_{\alpha\beta\gamma}\epsilon_{ij}\epsilon_{\delta \rho\gamma}+\lambda_8\tilde R_2^{\alpha i} S_1^{\beta*}\bar S_1^{\gamma*}H^{i*}\epsilon_{\alpha\beta\gamma}+\lambda_9\tilde R_2^{\alpha i} \tilde R_2^{\beta j}\tilde R_2^{\gamma k}H^{\ell*}\epsilon_{\alpha\beta\gamma}\epsilon_{ij}\epsilon_{k\ell} + {\rm h.c.}\ ,\label{eq-V4}
\end{align}
and where $\epsilon_{i_1\dots i_d}$ is the Levi-Civita tensor in $d$ dimensions (we use the convention $\epsilon_{1 2 \dots d}=+1$), $\sigma_{i}$ are the Pauli matrices and $t^a$ the Gell-Mann color matrices. Note how in particular the terms in Eq.~(\ref{eq-V3}) and  some of Eq.~(\ref{eq-V4})  (those involving two Higgs fields $H$) lead to cubic and quartic interactions, respectively, that provide contributions to the 1-loop gluon fusion diagrams for double Higgs production. In particular the cubic interactions, before EW symmetry breaking, contribute to the double cross section but not to the single one, showing a hint on the enhancement of double production in the presence of large cubic couplings. For the LHC phenomenology that we are interested in, we can neglect the quartic terms, except those with couplings: $\lambda_H, \lambda_{1,2,3}$ and $\lambda_1'$, where $\lambda_H$ is the quartic Higgs self-coupling. These terms, when the Higgs vacuum expectation value (vev) is considered, contribute to the LQ masses and they also can be relevant to single and double Higgs production.

It is clear from this potential that in principle all scalars involved could possibly obtain a vev in regions of the parameter space.  In order to have proper Electroweak symmetry breaking (EWSB) taking place and no color/EM charge breaking minima stemming from the LQs acquiring a vev, we will in the following show the conditions  that must be demanded to the couplings that enter in the potential~\cite{Camargo-Molina:2013sta,Chowdhury:2013dka,Blinov:2013fta}. 

Given the large amount of parameters that enter in the potential and the fact that we are interested in regions in which the LQs acquire vanishing vevs, we do an analysis of the extrema of the potential in which we neglect the quartic contributions, given that they will be only important in regions of large field values. We do require though that all of the quartic couplings are real, $\lambda_H$ and $\lambda_{1,2,3}$ positive, and the condition $\lambda_1 > |\lambda_1'|$ in order to guarantee a potential bounded from below. Thus the chances of obtaining a non-vanishing value for $\langle \tilde R_2 \rangle$,  $\langle  S_1 \rangle$ and $\langle \bar S_1 \rangle$ depend mostly on the parameters $\mu_1$, $\mu_2$, $m^2_{\tilde R_2}$, $m^2_{ S_1}$ and $m^2_{\bar S_1}$. The extrema condition then simplifies to
\begin{eqnarray}
\left.\frac{\delta \mathcal{L}}{\delta \tilde R_2}\right|_{\rm vev} &=&  \mu_1 \langle H \rangle^{\dagger}  \langle S_1 \rangle - \mu_2 \langle H \rangle^{T} \epsilon \langle \bar S_1 \rangle + m^2_{\tilde R_2} \langle \tilde{R_2} \rangle^{\dagger}=0 \, ,\\
\left.\frac{\delta \mathcal{L}}{\delta S_1}\right|_{\rm vev} &=&  \mu_1 \langle H \rangle^{\dagger}  \langle \tilde R_2 \rangle +m^2_{ S_1} \langle S_1 \rangle^{*}=0 \, ,\\
\left.\frac{\delta \mathcal{L}}{\delta \bar S_1}\right|_{\rm vev} &=& - \mu_2 \langle H \rangle^{T} \epsilon \langle \tilde R_2 \rangle + m^2_{\bar S_1} \langle \bar{S_1} \rangle^{*}=0 \, ,
\end{eqnarray}
which in the unitary gauge for $H$ and separating the up and down components of $\tilde R_2$ leads to
\begin{eqnarray}
\mu_1\frac{v}{\sqrt{2}}\langle S_1 \rangle^{*}+ m^2_{\tilde R_2} \langle \tilde{R}_2^{d}\rangle&=&0 \, , \nonumber\\
\mu_1\frac{v}{\sqrt{2}}\langle \tilde{R}_2^{d} \rangle+ m^2_{S_1} \langle S_{1}\rangle^{*}&=&0 \, ,\label{ext1}
\end{eqnarray}
\begin{eqnarray}
\mu_2\frac{v}{\sqrt{2}}\langle \bar{S}_1 \rangle^{*}+ m^2_{\tilde R_2} \langle \tilde{R}_2^{u}\rangle&=&0 \, , \nonumber\\
\mu_2\frac{v}{\sqrt{2}}\langle \tilde{R}_2^{u} \rangle+ m^2_{\bar{S}_1} \langle \bar{S}_{1}\rangle^{*}&=&0 \, .\label{ext2}
\end{eqnarray}
The system described in Eqs.~(\ref{ext1}) and~(\ref{ext2}) have a vanishing solution for the LQs vevs whenever
\begin{eqnarray}
\mu_1^2\frac{v^2}{2}&\neq& m^2_{\tilde R_2}m^2_{S_1} \, , \nonumber\\
\mu_2^2\frac{v^2}{2}&\neq& m^2_{\tilde R_2}m^2_{\bar S_1} \, .
\end{eqnarray}
As long as we consider values in which this is satisfied, the LQs will not acquired a color/EM charged breaking vev. This analysis does not guarantee that another vev for the LQs may develop at other values of the fields, it only assures that the EWSB vev $v=246$ GeV with $ \langle \tilde{R_2} \rangle =  \langle S_1 \rangle = \langle \bar S_1 \rangle=0$ is an extremum of the potential. Demanding that all masses for the physical states are positive provides alongside the minimum condition for vevs just mentioned. The possibility for the LQs acquiring a vev at large values of the fields can be somewhat lessen in the case of sizable positive quartic couplings and though we do not perform a thorough analysis of color/EM charge breaking minima and the stability (or metastability) of the EWSB vacuum, our results for the double Higgs production do not strongly depend on the quartic coupling values. 
After EWSB  takes place, once the SM Higgs gets a vev, there is mixing among the different LQs components (the upper component of the doublet, $\tilde R_2^{u}$ with EM charge $Q=2/3$, mixes with $\bar S_1$, and the lower component $\tilde R_2^{d}$ with EM charge $Q=-1/3$ mixes with $S_1$). This can be seen, once EWSB takes place, by writing the mass matrices involving $\tilde R_2^{d}$,  $S^*_1$, and $\tilde R_2^{u}$, $\bar S^*_1$ 
\begin{eqnarray}\label{M2d}
\mathcal{M}^2_{d}=\left(
\begin{array}{cc}
m_{\tilde{R}_2}^2  + \frac{v^2}{2}(\lambda_1-\lambda^{\prime}_1) & \mu_1 \frac{v}{\sqrt{2}} \\
 \mu_1^* \frac{v}{\sqrt{2}} & m_{S_1}^2 +\lambda_2  \frac{v^2}{2}
\end{array}
\right),
\end{eqnarray}
\begin{eqnarray}\label{M2u}
\mathcal{M}^2_{u}=\left(
\begin{array}{cc}
m_{\tilde{R}_2}^2  + \frac{v^2}{2}(\lambda_1+\lambda^{\prime}_1) & \mu_2 \frac{v}{\sqrt{2}} \\
 \mu_2^* \frac{v}{\sqrt{2}} & m_{\bar S_1}^2 +\lambda_3  \frac{v^2}{2}
\end{array}
\right),
\end{eqnarray}
These mass matrices have to be simultaneously diagonalized. We assume that all parameters are real. We can find the eigenvalues and eigenvectors of both mass matrices which represent the physical fields analytically
\begin{eqnarray} \label{eq-rot1}
\left(
\begin{array}{cc}
\chi^d_2  \\
\chi^d_1
\end{array}
\right)
=
\begin{array}{c}
U_d
\end{array}
\left(
\begin{array}{cc}
\tilde{R}_2^{d} \\
S_{1}^{*}
\end{array}
\right)
=
\left(
\begin{array}{ccc}
-\sin\theta_{d}  & \cos\theta_{d}  \\
\cos\theta_{d}  & \sin\theta_{d} 
\end{array}
\right)
\left(
\begin{array}{cc}
\tilde{R}_2^{d} \\
S_{1}^{*}
\end{array}
\right)~,
\end{eqnarray}
\begin{eqnarray} \label{eq-rot2}
\left(
\begin{array}{cc}
\chi^u_2  \\
\chi^u_1
\end{array}
\right)
=
\begin{array}{c}
U_u
\end{array}
\left(
\begin{array}{cc}
\tilde{R}_2^{u} \\
\bar{S}_{1}^{*}
\end{array}
\right)
=
\left(
\begin{array}{ccc}
-\sin\theta_{u}  & \cos\theta_{u}  \\
\cos\theta_{u}  & \sin\theta_{u} 
\end{array}
\right)
\left(
\begin{array}{cc}
\tilde{R}_2^{u} \\
\bar{S}_{1}^{*}
\end{array}
\right)~.
\end{eqnarray}
with the eigenvalues as
\begin{align}
m^{2}_{\chi^d_{2,1}}=\frac{1}{2}\left(m^2_{\tilde{R}_2}+m^2_{S_1}+\frac{v^2}{2}(\lambda_1-\lambda^{\prime}_1+\lambda_2)\mp\sqrt{\left( m^2_{\tilde{R}_2}-m^2_{S_1} +\frac{v^2}{2}(\lambda_1-\lambda^{\prime}_1-\lambda_2) \right)^2+2\mu_1^2 v^2}\right) \nonumber\\
m^{2}_{\chi^u_{2,1}}=\frac{1}{2}\left(m^2_{\tilde{R}_2}+m^2_{\bar S_1}+\frac{v^2}{2}(\lambda_1+\lambda^{\prime}_1+\lambda_3)\mp\sqrt{\left( m^2_{\tilde{R}_2}-m^2_{\bar S_1} +\frac{v^2}{2}(\lambda_1+\lambda^{\prime}_1-\lambda_3) \right)^2+2\mu_2^2 v^2}\right)\nonumber\\
\end{align}
A simple relationship can be found between the mixing angles and the mass eigenvalues
\begin{equation}
2\mu_1^2 v^2=(\sin2\theta_{d})^2(m^2_{\chi^d_1}-m^2_{\chi^d_2})^2,~\quad 2\mu_2^2 v^2=(\sin2\theta_{u})^2(m^2_{\chi^u_1}-m^2_{\chi^u_2})^2
\end{equation}
Through these sets of equations and from a practical point of view, we are able to write the Lagrangian parameters in terms of the physical mass eigenvalues and mixing angles. This turns out to be very helpful when calculating numerically the Higgs production at the LHC. In Appendix~\ref{sec:appendix:Higgs-leptoquarksphysicalbasis} we derive the couplings between the Higgs and LQs in the physical basis, and in Appendix~\ref{perturbativeexpansion} we present a Naive Dimensional Analysis (NDA) of the size of the couplings to be under control in  the perturbative expansions.

Finally, we write the Lagrangian for the LQs and the SM fermions in the interaction basis. In the absence of $\nu_R$
\begin{align}\label{eq-ferm-int}
& {\cal L}_{\tilde R_2}= y_{2\ell}\ \bar d_R \tilde R_2 \epsilon \ell_L +{\rm h.c.} \ ,\nonumber \\
& {\cal L}_{\bar S_1}= z_{\bar S_1d}\ \bar d_R^c \bar S_1^* d_R +{\rm h.c.} \ ,\nonumber \\
& {\cal L}_{S_1}= y_{S_1\ell}\ \bar q_L^c S_1 \epsilon \ell_L +y_{S_1e}\ \bar u_R^c S_1 e_R + z_{S_1 q_L}\ \bar q_L^c S_1^{*} \epsilon q_L + z_{S_1 q_R}\ \bar u_R^c S_1^{*} d_R+ {\rm h.c.} \ . \end{align}
We have not included generation indices and the $z$-coupling is antisymmetric. The presence of $\nu_R$ would introduce new interactions, opening more decay channels.

\section{Higgs phenomenology}
\label{higgs phenomenology}

The discovery of a scalar particle compatible with the Standard Model Higgs boson at the Large Hadron Collider (LHC) has motivated searches for new physics using the Higgs boson as a probe. In particular, Higgs pair production is not only interesting as a probe of the trilinear Higgs self-coupling, but because its rate could be sensible to the presence of new physics effects.

Several mechanisms give rise to the production of pairs of neutral Higgs bosons in hadrons collisions. Multiple Higgs states can be excited through Higgs-strahlung off $W$ bosons or through $WW$ fusion in proton-proton collisions but, the largest production rates are provided by gluon fusion. Since Higgs bosons are colorless particles, their pair production from the collision of two gluons is necessarily mediated through loop contributions which could be permeable to new physics effects. In this section, we explore the Higgs production in gluon fusion under the influence of scalar LQs. First, we focus on the dissection of the ultraviolet complete calculation of the double Higgs production cross section, at leading order on $\alpha_\text{S}$. Then, we review it but from the perspective of the Higgs LET framework, comparing both results to check the limit of this approximation. Finally, we take advantage of the simplicity of the results obtained through the LET and evaluate the correlation between the double Higgs production cross section and the single Higgs production, which is certainly more constrained by the experimental data. We also consider Higgs decays into a pair of photons and into a $Z$ boson and a photon, where in both cases the LQs contribute at one loop level. However, we observe that their impact on the phenomenology is mild.

All the results shown throughout the paper are given at leading order. Notice, however, that when considering the phenomenological impact of the double Higgs production cross section in Secs.~\ref{sec-light-scenario} and~\ref{sec-heavy-scenario} we take ratios against the SM prediction at leading order. Therefore, we expect that most high order corrections (in particular QCD ones) drop in the ratio.\footnote{Higher order QCD corrections to the double Higgs production have been performed under different approximations, for example in~\cite{deFlorian:2013uza,deFlorian:2013jea,Maltoni:2014eza,Frederix:2014hta,Grigo:2014jma,Grigo:2015dia,Borowka:2016ehy,Degrassi:2016vss,deFlorian:2016uhr,Spira:2016zna,Borowka:2016ypz,Heinrich:2017kxx,Jones:2017giv,Banerjee:2018lfq,Baglio:2018lrj,Chen:2019lzz,Chen:2019fhs,Baglio:2020ini}.}

\subsection{Double Higgs production}

The Lorentz invariant differential cross section of gluon fusion into Higgs pairs could be split into two main pieces as follows~\cite{Glover:1987nx,Plehn:1996wb,Dawson:1998py,BarrientosBendezu:2001di}

\begin{equation}
    \frac{\text{d}\hat{\sigma}_{gg}}{\text{d}\hat{t}} = \frac{G_F^2~ \alpha_\text{S}^2}{2^{14} \pi^3~ \hat{s}^2}~~ \Big[~ \big|\mathcal{F}\big|^2 + \big|\,\mathcal{G}\,\big|^2 \Big]
    \label{eq:ggHH-diffxs} \, ,
\end{equation}
where $\hat{s}$ represents the invariant mass of the Higgs pair finally produced and $\hat{t}$ is the transferred squared momentum from one of the gluons in the initial state to one of the Higgs bosons states. The form factors, $\mathcal{F}$ and $\mathcal{G}$, hold the contributions linked to the same or the opposite polarization configurations of the incoming gluons, respectively.

\begin{figure}[h!]
    \centering
    \includegraphics[width=0.3\textwidth]{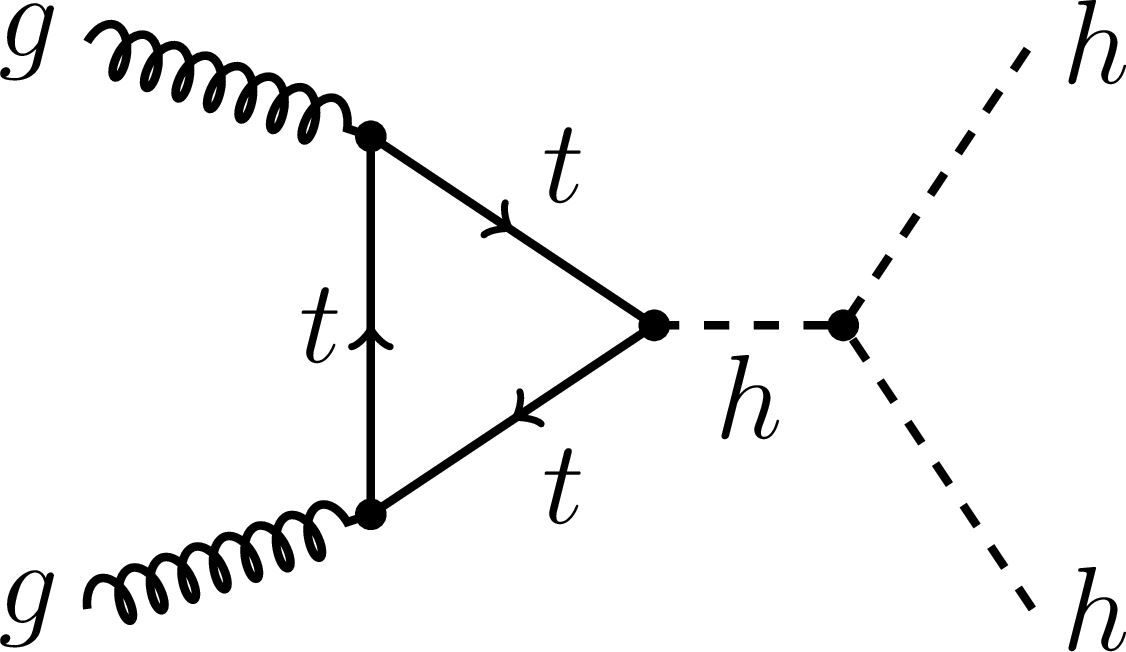}\hspace*{2.cm}
    \includegraphics[width=0.3\textwidth]{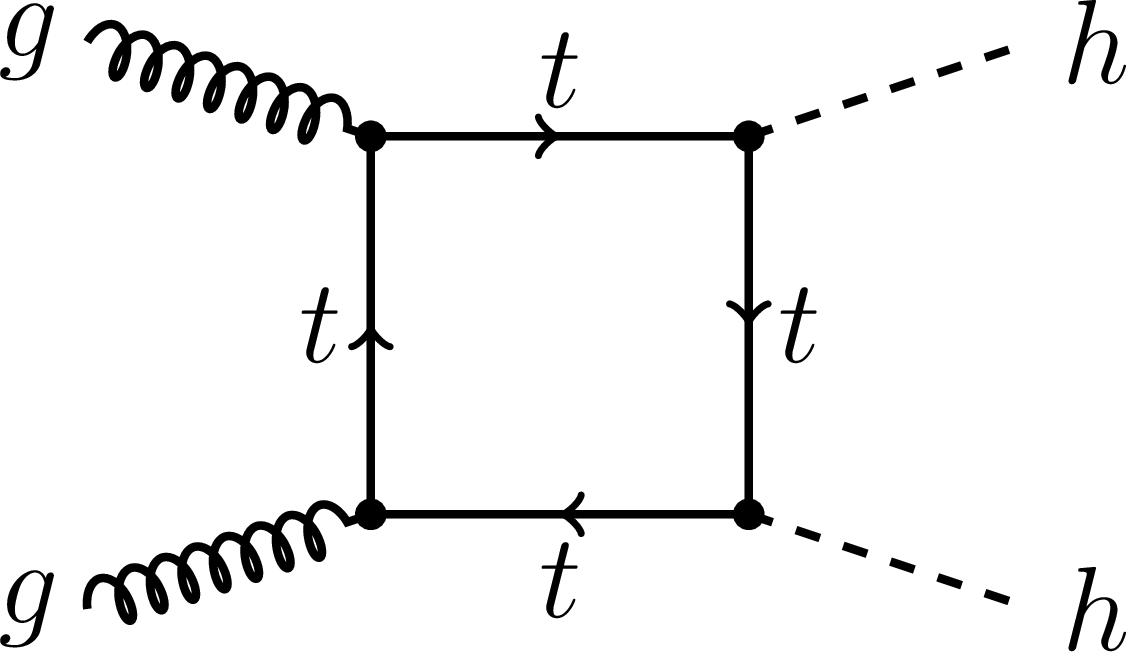}
    \caption{Generic diagrams for the triangle and box contributions to the pair production of Higgs bosons in gluon-gluon collisions.}
    \label{SM}
\end{figure}

\begin{figure}[h!]
    \centering
    \begin{subfigure}[b]{\textwidth}
        \centering
        \includegraphics[width=.3\textwidth]{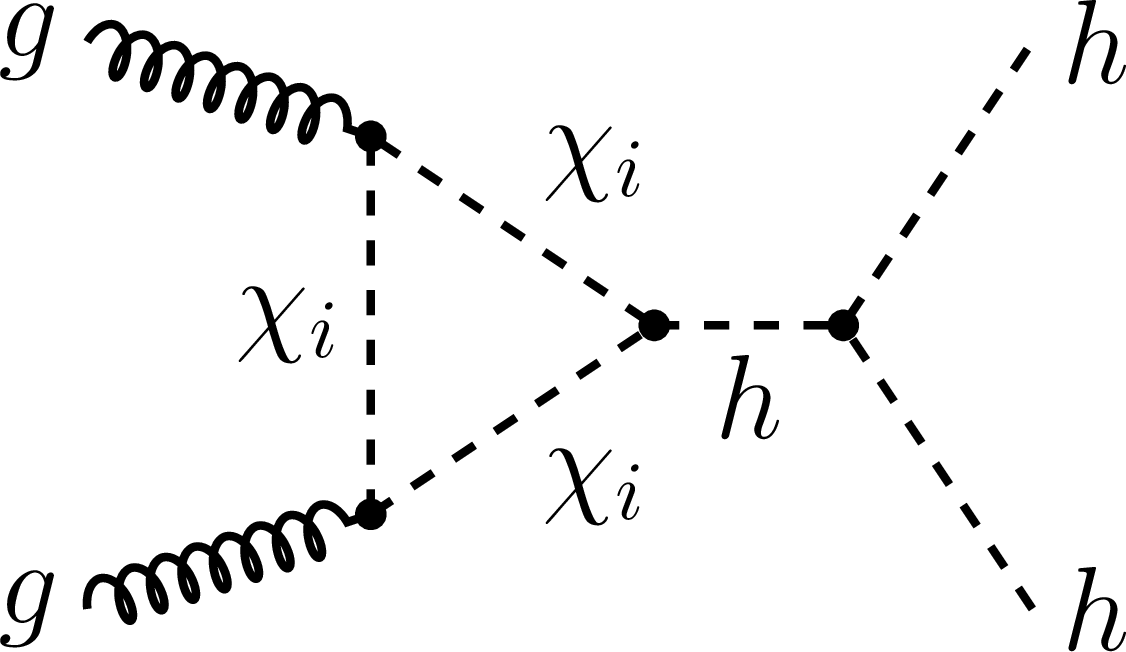}\hspace*{2.cm}
        \includegraphics[width=.3\textwidth]{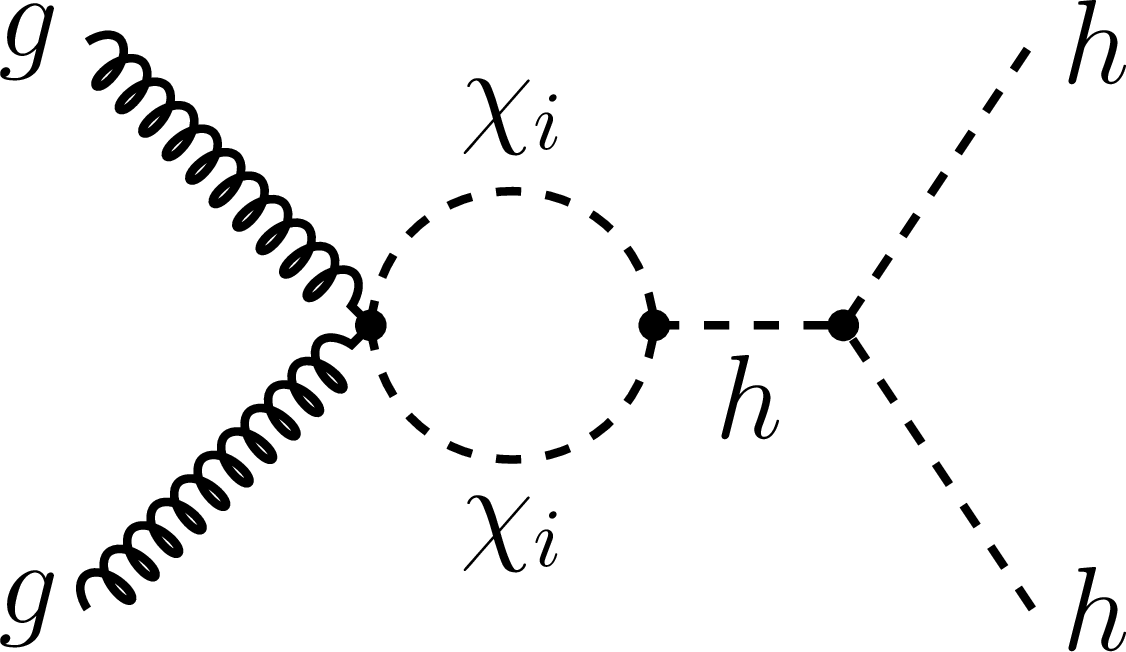}
        \caption{Contributions to $\mathcal{F}^{(1)}_{ij}$ and $\mathcal{G}^{(1)}_{ij}$ form factors.}
    \end{subfigure}
 
    \vspace{1.2cm}
     \centering
    \begin{subfigure}[b]{\textwidth}\hspace*{2.2cm}
        \includegraphics[width=.3\textwidth]{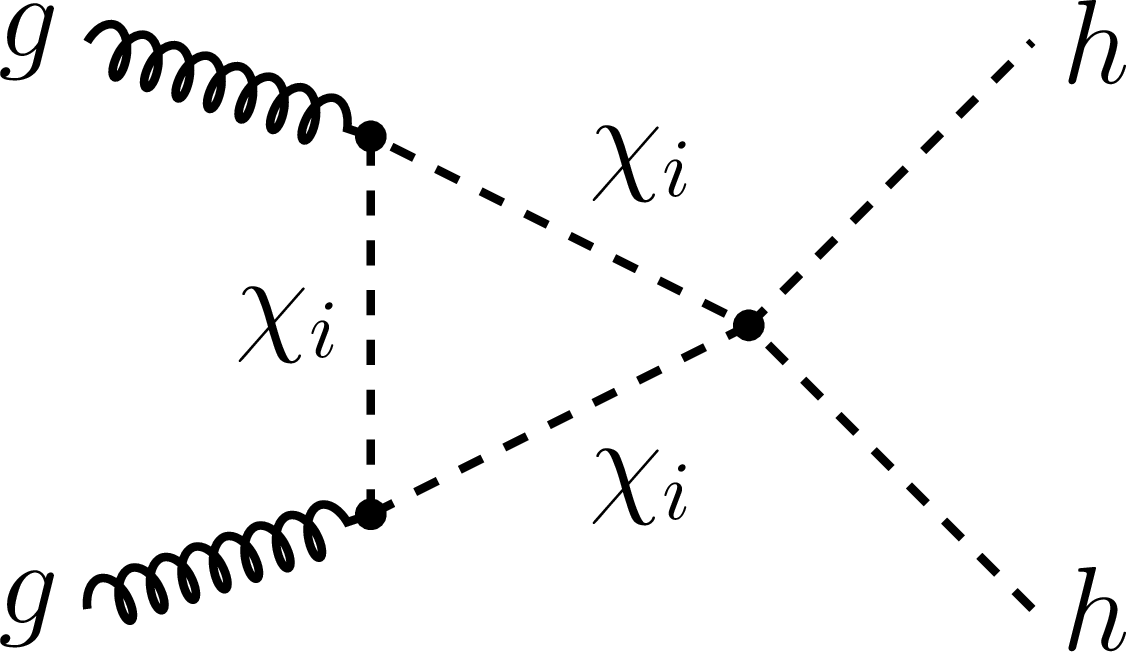}\hspace*{2.cm}
        \includegraphics[width=.3\textwidth]{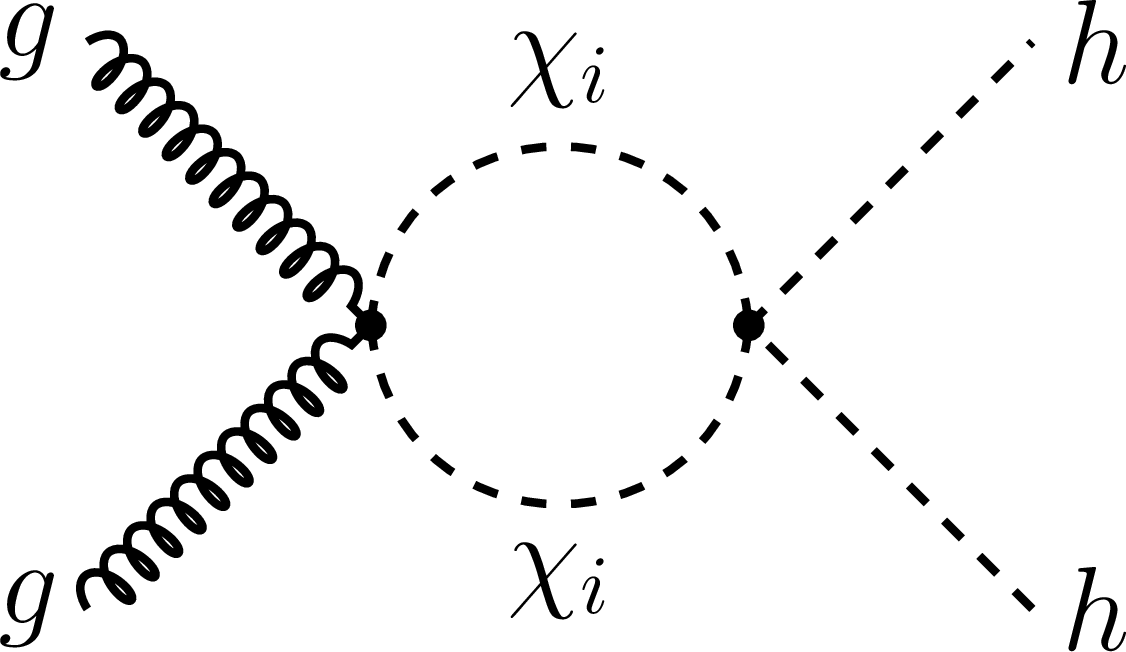}
        \caption{Contributions to $\mathcal{F}^{(2)}_{ij}$ and $\mathcal{G}^{(2)}_{ij}$ form factors.}
    \end{subfigure}

    \vspace{1.2cm}
    \begin{subfigure}[b]{0.35\textwidth}
        \centering
        \includegraphics[width=.851\textwidth]{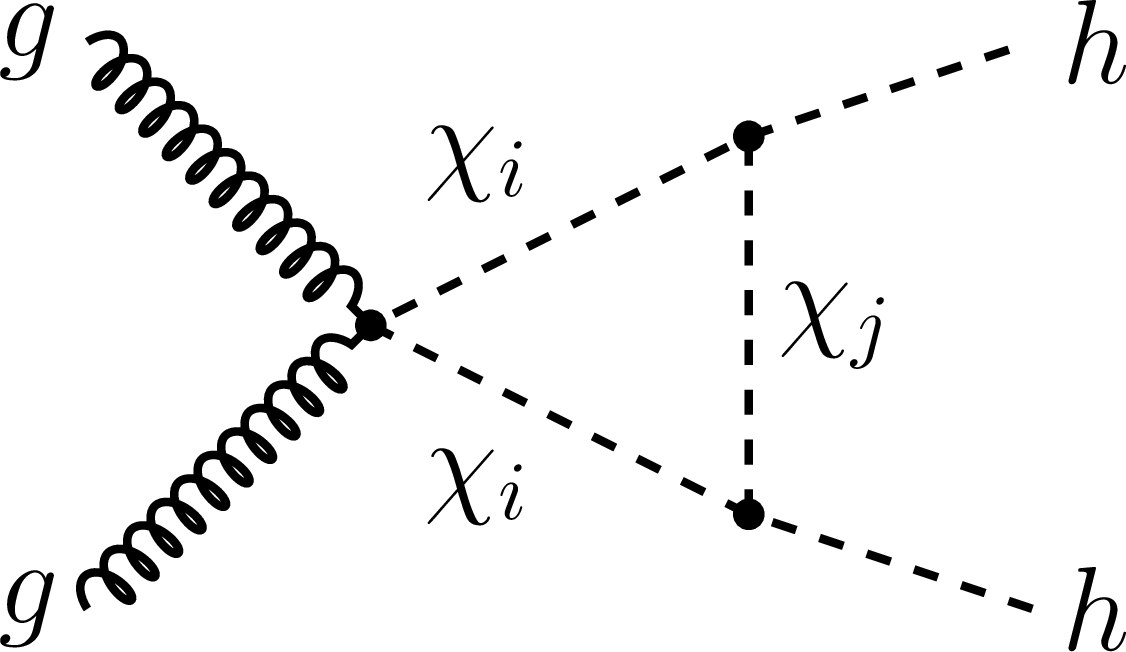}
        \caption{Contributions to $\mathcal{F}^{(3)}_{ij}$ and $\mathcal{G}^{(3)}_{ij}$ form factors.}
        \label{fig:lqrk_diagrams:d}
    \end{subfigure}
    \hspace*{1cm}
    \begin{subfigure}[b]{0.35\textwidth}
        \centering
        \includegraphics[width=.851\textwidth]{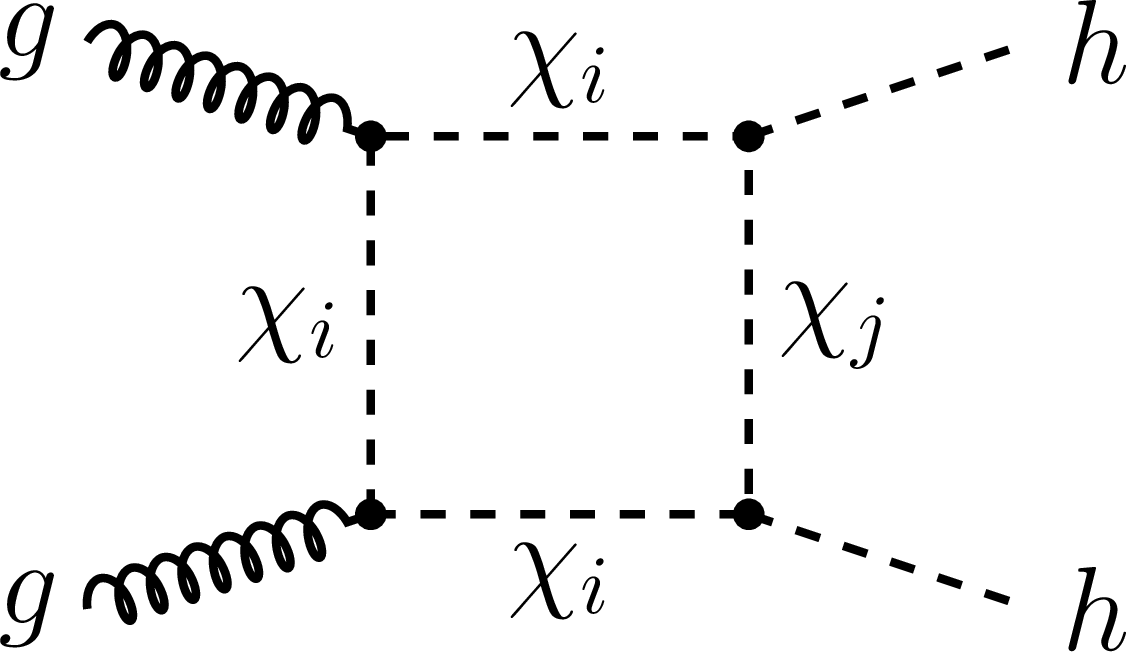}
        \caption{Contributions to $\mathcal{F}^{(4)}_{ij}$ and $\mathcal{G}^{(4)}_{ij}$ form factors.}
        \label{fig:lqrk_diagrams:e}
    \end{subfigure}
        \caption{Generic diagrams that contribute to the new physics form factors of gluon fusion into Higgs pairs, mediated by scalar LQs.}
        \label{fig:lqrk_diagrams}
\end{figure}

Each of these form factors could be generically written as the sum of several terms related with various loop configurations
\begin{align}
    \mathcal{F} = \mathcal{F}_\triangle + \mathcal{F}_\square + \sum_{k=u,d} \bigg\{ &\sum_{i=1,2} \Big[ \mathcal{F}^{(1)}_{\chi^k_i}+ \mathcal{F}^{(2)}_{\chi^k_i} \Big]+\frac{1}{2}\sum_{i=1,2}\sum_{j=1,2} \Big[ \mathcal{F}^{(3)}_{\chi^k_i\chi^k_j} + \mathcal{F}^{(4)}_{\chi^k_i\chi^k_j} \Big] \bigg\} \, ,
    \label{eq:fFF}
\end{align}
\begin{align}
    \mathcal{G} = \mathcal{G}_\triangle + \mathcal{G}_\square + \sum_{k=u,d} \bigg\{ &\sum_{i=1,2} \Big[ \mathcal{G}^{(1)}_{\chi^k_i}+ \mathcal{G}^{(2)}_{\chi^k_i} \Big]+\frac{1}{2}\sum_{i=1,2}\sum_{j=1,2} \Big[ \mathcal{G}^{(3)}_{\chi^k_i\chi^k_j} + \mathcal{G}^{(4)}_{\chi^k_i\chi^k_j} \Big] \bigg\} \, .
    \label{eq:gFF}
\end{align}
The first two terms of each form factor are provided by the Standard Model, through the contributions mediated by triangle and box loops of a top quark. Taking into account that this form factors are proportional to the mass of the quark involved in the loop, the remaining flavors bring negligible contributions. Representative diagrams are shown in Fig.~\ref{SM}. The remaining terms arise from the loops induced by the presence of the mass eigenstate LQs, $\chi^u_1$, $\chi^u_2$, $\chi^d_1$ and $\chi^d_2$, presented in Sec.~\ref{model}, and their interaction with the Higgs boson. The generic diagrams that contribute to these terms are represented in Fig.~\ref{fig:lqrk_diagrams}. It is worth mentioning that the new trilinear effective couplings  enable the possibility of blending different types of physical LQs in the loops, see Fig.~\ref{fig:lqrk_diagrams:d} and \ref{fig:lqrk_diagrams:e}. This feature is not present in the Standard Model diagrams due to the diagonal nature of the Higgs couplings to quarks. The dependence of $\mathcal{F}_\triangle$, $\mathcal{F}_\square$, $\mathcal{G}_\triangle$, $\mathcal{G}_\square$ and the new physics form factors on the Mandelstam variables $\hat{s}$, $\hat{t}$ and $\hat{u}$ can be found in the Appendix \ref{sec:appendix:formfactos}. 

\begin{figure}[h!]
    \centering
    \includegraphics[width=0.8\textwidth]{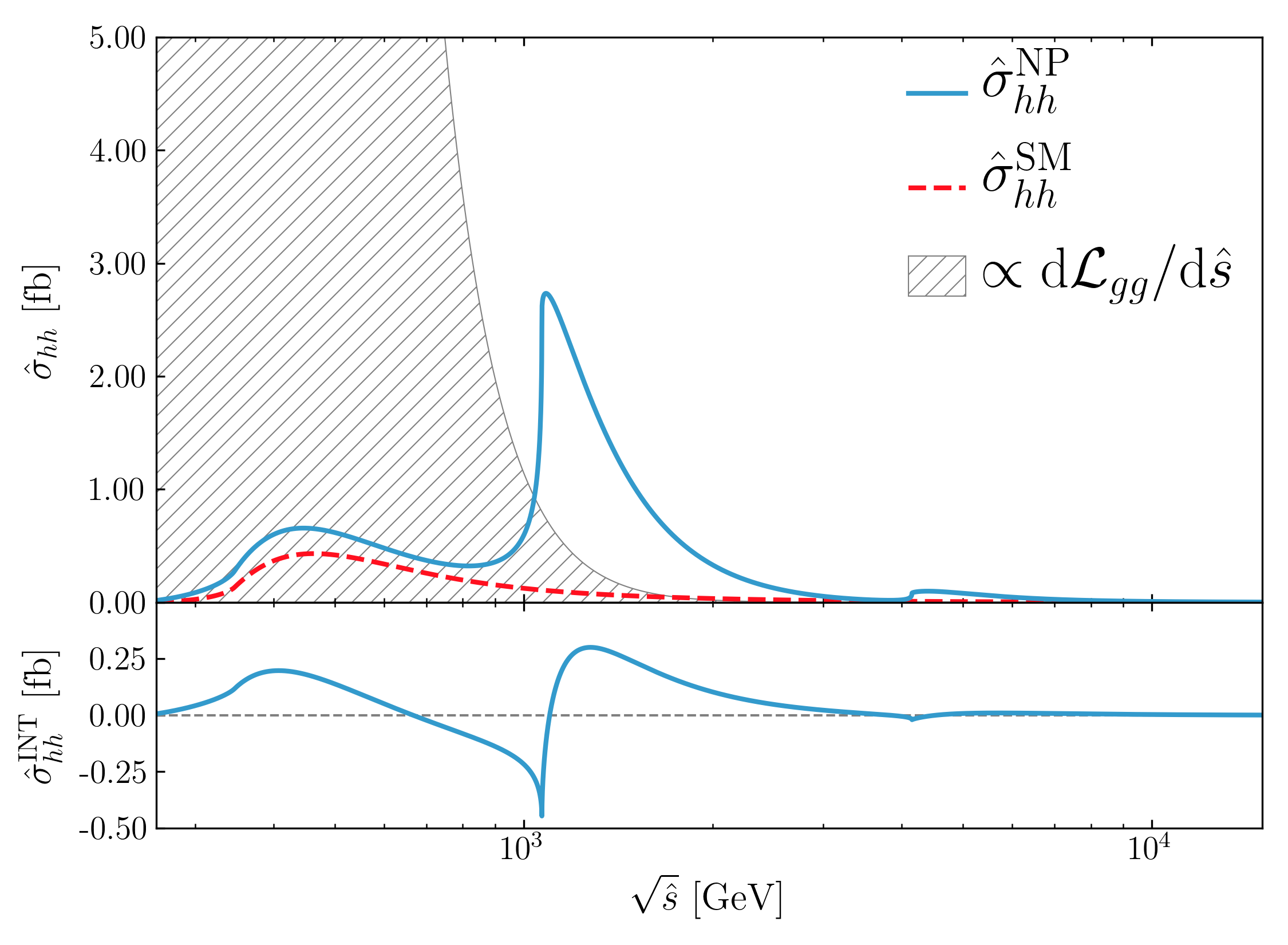}
    \caption{In the upper panel we have the Standard Model contribution (red dashed line) to the parton cross section of di-Higgs production in gluon fusion as a function of the center-of-mass energy of the gluon pair. Its modification by the presence of LQs (with masses of 535 GeV and 2076 GeV) is represented by the blue solid line. The shaded area is proportional to the gluon-gluon luminosity. In the lower panel we have the interference between the Standard Model and the LQ form factors.}
    \label{fig:ggHH_vs_s}
\end{figure}

In the upper frame of Fig.~\ref{fig:ggHH_vs_s}, we can see how the differential cross section of Eq.~(\ref{eq:ggHH-diffxs}) looks as a function of the center-of-mass energy of the gluon system $\sqrt{\hat{s}}$, after being integrated in $\hat{t}$. The red dashed line represents the purely Standard Model result. At the leading order, the shape of the cross section is basically dominated by the squared module of the box diagrams and the destructive interference between the former and the triangle diagram. The blue solid line is produced by the inclusion of a pair of scalar LQs in the loop, obtained from the diagonalization of the mass matrices of Eqs.~(\ref{M2d}) and~(\ref{M2u}) with arbitrary parameters that satisfy all the experimental constraints considered in this work. The Standard Model cross section is reshaped by the presence of two resonance peaks at twice the masses of the LQs, taken around 0.5 TeV and 2 TeV. The interference between the new physics form factors and the Standard Model contribution is responsible for the final shape of the resonance peaks and the enhancement of the cross-section at low values of $\hat{s}$, see the lower panel of Fig.~\ref{fig:ggHH_vs_s}. As it can also be appreciated in this figure, the gluon-gluon luminosity ${\text{d}\mathcal{L}_{gg}}/{\text{d}\hat{s}}$ increases extremely fast as $\hat{s}$ decreases. Consequently, the enhancement produced by the interference between the Standard Model and the new physics contributions results especially relevant to boost the double Higgs production in proton-proton collisions.

The total cross section for Higgs pair production can be derived by integrating over $\hat{t}$ and performing the convolution of the parton cross section with the gluon-gluon luminosity as follows

\begin{equation}
    \sigma_{hh}(s) = \int^{1}_{4m_h^2 / s}d\tau~ \frac{\text{d}\mathcal{L}_{gg}}{\text{d}\tau}(\tau,s) \times \int^{t_+(\tau s)}_{t_-(\tau s)}d\hat{t}~ \frac{\text{d}\hat{\sigma}_{hh}}{\text{d}\hat{t}}(\tau s,\hat{t}),
\end{equation}
where $s$ denotes the center-of-mass energy of the proton to proton collision that gives rise to the gluon fusion process and where we made use of the relation $\hat{s}=\tau s$. The limits of integration are defined as $t_\pm(\hat{s}) = {1}/{2} \times [2 m_h^2 - \hat{s} \pm \sqrt{\hat{s}^2 - 4m_h^2 \hat{s}}]$. To compute the gluon luminosity ${\text{d}\mathcal{L}_{gg}}/{\text{d}\tau}$ we made use of the set of parton distribution functions (PDF) PDF4LHC15\_nnlo\_mc, provided by the PDF4LHC working group~\cite{Butterworth:2015oua} at next-to-next-to-leading order.

\subsection{Higgs low energy theorem}\label{LET}
The LET allows to obtain the interactions with photons and gluons generated at loop level by heavy particles, in the limit where their masses are large compared with the other scales~\cite{Ellis:1975ap,Shifman:1979eb}. We are interested in the couplings of one and two Higgs with two gluons, involved in single and double Higgs production. In the following we include the top quark within the heavy particles, as well as the LQs, although it is well known that the heavy top mass limit disagrees with the full calculation by 20$\%$ for di-Higgs production. Similar discussions of LETs with colored scalars in the context of supersymmetry have been considered, for example, in Ref.~\cite{Wagner}.

For the interactions of the Higgs with gluons we use the following effective Lagrangian
\begin{equation}
{\cal L}_{\rm eff} \supset -\frac{\alpha_\text{S}}{32\pi}G^a_{\mu\nu}G^{a\mu\nu}\sum_i\beta_i \log\frac{\Lambda^2}{m_i^2(v)} \, .
\end{equation}
Replacing $v\to v+h$ and expanding $m_i$ in powers of $h$
\begin{equation}
{\cal L}_{\rm eff}  = \frac{\alpha_\text{S}}{32\pi}G^a_{\mu\nu}G^{a\mu\nu} \left[c_0 + g_h \frac{h}{v} + \frac{g_{hh}}{2} \left(\frac{h}{v}\right)^2 + \dots \right] \, ,
\end{equation}
with
\begin{align}
g_h& =v \sum_i \beta_i \partial_v \log m_i^2(v) = 
\sum_i \beta_i \frac{\partial [\log m_i^2(v)]}{\partial(\log v)} = 
\sum_{\rm s} \beta_s \frac{\partial [\log {\rm det} \mathcal{M}_s^2(v)]}{\partial(\log v)} 
,
\label{eff-gh}\\
g_{hh}& = v^2 \sum_i \beta_i \partial_v^2 \log m_i^2(v) = 
\sum_i \beta_i \left[\frac{\partial^2 [\log m_i^2(v)]}{\partial(\log v)^2}-\frac{\partial [\log m_i^2(v)]}{\partial(\log v)}\right] 
 \nonumber \\
& = 
\sum_{\rm s} \beta_s \left[\frac{\partial^2 [\log {\rm det} \mathcal{M}_s^2(v)]}{\partial(\log v)^2} - \frac{\partial [\log {\rm det} \mathcal{M}_s^2(v)]}{\partial(\log v)}\right] \ , 
\label{eff-ghh}
\end{align}
where $i$ runs over all the particles, $\beta_i=2/3, 1/6$ for Dirac fermions and complex scalars, respectively, are the coefficients of the $\beta$ function of QCD, and $s$ labels the species of particles: up-type LQs, down-type LQs and the top (we neglect the effect of the other SM fields).

For an eigenmass LQ $\chi$, the coupling $|\chi|^2h$ can be obtained from: $\mathscr{C}_{\chi\chi}=\partial_v m_\chi^2(v)$. Thus
\begin{equation}\label{ghLET}
g_h = \beta_t g^t_h + v\beta_{\rm LQ}\sum_\chi \frac{\mathscr{C}_{\chi\chi}}{m_\chi^2} \ ,
\end{equation}
where $g^t_h$ arises from the top quark contribution: $g^t_h = 2$.\footnote{Using the proper values of $\beta_i$, as well as $1/8\simeq 0.12$, Eq.~(\ref{ghLET}) leads to the same result as the one of Eq.~(\ref{deltakappag}) in the limit of large top and LQ masses, that gives a very good approximation for single Higgs production.}

For our model with LQs, we split the contributions of the up- and down-type states
\begin{equation}
g_h = \beta_t g^t_h \left(1+\frac{g^{\rm LQu}_h+g^{\rm LQd}_h}{4g^t_h}\right)
\end{equation}
with
\begin{equation}
g^{\rm LQ}_h = \frac{\partial [\log {\rm det} \mathcal{M}_{\rm LQ}^2(v)]}{\partial(\log v)} \ .
\end{equation}
These effective couplings can be computed straightforwardly from the mass matrix. Thus, we have a simple explicit expression for these couplings in terms of the parameters of the model
\begin{align}
& g_h^{\rm LQu}=\frac{4 v^2 \left[-\mu_2^2+\lambda_3 m_{\tilde{R}_2}^2+(\lambda_1+\lambda_1') \left(m_{\bar S_1}^2+\lambda_3 v^2\right)\right]}{\left(2 m_{\bar S_1}^2+\lambda_3 v^2\right) \left(2 m_{\tilde{R}_2}^2+v^2 (\lambda_1+\lambda_1')\right)-2 \mu_2^2 v^2} \ ,
\\
& g_h^{\rm LQd}=\frac{4 v^2 \left[-\mu_1^2+\lambda_2 m_{\tilde{R}_2}^2+(\lambda_1-\lambda_1') \left(m_{S_1}^2+\lambda_2 v^2\right)\right]}{\left(2 m_{S_1}^2+\lambda_2 v^2\right) \left(2 m_{\tilde{R}_2}^2+v^2 (\lambda_1-\lambda_1')\right)-2\mu_1^2 v^2} \ .
\end{align}

\noindent Similar Eqs. can be obtained for the contributions to $g_{hh}$: $g_{hh}^{\rm LQu}$ and $g_{hh}^{\rm LQd}$, by using Eq.~(\ref{eff-ghh}). 

It results quite illuminating to rewrite the expressions for $g_h$ and $g_{hh}$ in terms of the physical masses. For that purpose it is useful to consider that $g_h =\sum_i \beta_i \frac{\partial [\log m_i^2(v)]}{\partial(\log v)}$ and that $g_{hh}=v\partial_{v}g_h -g_h$. 

Considering only the LQ contributions, concentrating initially in the calculation of $g_h$ and using the expressions for the LQ physical masses, it is easy to calculate how the cubic and quartic interactions contribute from the up and down sectors
\begin{align}
g_{h,\; cubic}^{\rm u}&=-\frac{\beta_{\rm LQ}v^2\mu^2_2}{m^2_{\chi^{u}_1} m^2_{\chi^{u}_2}}\label{ghcubicup}
\\
g_{h,\; quartic}^{\rm u}&= \frac{\beta_{\rm LQ}v^2}{2m^2_{\chi^{u}_1} m^2_{\chi^{u}_2}}\bigg[ 
(\lambda_1+\lambda'_{1}+\lambda_3)(m^2_{\chi^{u}_1}+m^2_{\chi^{u}_2}) \nonumber \\
& - (\lambda_1+\lambda'_{1}-\lambda_3)\sqrt{(m^2_{\chi^{u}_1}-m^2_{\chi^{u}_2})^2-2\mu^2_2 v^2}  \bigg]
\\
g_{h,\; cubic}^{\rm d}&=-\frac{\beta_{\rm LQ}v^2\mu^2_1}{m^2_{\chi^{d}_1} m^2_{\chi^{d}_2}}
\\
g_{h,\; quartic}^{\rm d}&= \frac{\beta_{\rm LQ}v^2}{2m^2_{\chi^{d}_1} m^2_{\chi^{d}_2}}\bigg[ (\lambda_1-\lambda'_{1}+\lambda_2)(m^2_{\chi^{d}_1}+m^2_{\chi^{d}_2}) \nonumber \\
& - (\lambda_1-\lambda'_{1}-\lambda_2)\sqrt{(m^2_{\chi^{d}_1}-m^2_{\chi^{d}_2})^2-2\mu^2_1 v^2}  \bigg]
\end{align}
where $\beta_{\rm LQ}$ $g^{\rm LQ}_h=g_{h,\;cubic}^{\rm u}+g_{h,\;quartic}^{\rm u}+g_{h,\;cubic}^{\rm d}+g_{h,\;quartic}^{\rm d}$. Notice how the second term inside the parenthesis in both $g_{h,\;quartic}^{\rm u}$ and $g_{h,\;quartic}^{\rm d}$ tend to be smaller than the first term. Comparing against the sign from the cubic contributions, this implies that the quartic tend to contribute with opposite sign. This opposing behavior is seen even in the numerical scans in the full UV theory. Furthermore, for $\mu^2_2\gg m^2_{\chi^{u}_1}$ and/or $\mu^2_1\gg m^2_{\chi^{d}_1}$, the cubic contributions dominate over the quartic ones, $g_{h,\;cubic}\gg g_{h,\;quartic}$.

In order to obtain a simple expression and assuming that the cubic terms dominate, we can also calculate the contributions to $g^{\rm LQ}_{hh}$
\begin{eqnarray}
g_{hh,\; cubic}^{\rm u}&=&-\frac{\beta_{\rm LQ}v^2\mu^2_2}{m^2_{\chi^{u}_1} m^2_{\chi^{u}_2}}-\frac{\beta_{\rm LQ}v^4\mu^4_2}{m^4_{\chi^{u}_1} m^4_{\chi^{u}_2}} \, , \label{ghhcubicup} \\
g_{hh,\; cubic}^{\rm d}&=&-\frac{\beta_{\rm LQ}v^2\mu^2_1}{m^2_{\chi^{d}_1} m^2_{\chi^{d}_2}}-\frac{\beta_{\rm LQ}v^4\mu^4_1}{m^4_{\chi^{d}_1} m^4_{\chi^{d}_2}} \, .
\end{eqnarray}

\subsubsection{Double Higgs production}
The calculation of double Higgs production cross section using LETs is greatly simplified. The amplitude is simply given by the sum of two terms, one with the $hhgg$ vertex, and another one with the $hgg$ vertex, a Higgs propagator and the Higgs trilinear coupling. To each kind of term there are contributions from the top and LQs. The hadronic cross section is~\cite{1206.7120}
\begin{equation}
\label{eq-xsec-eff}
\sigma_{hh}=\int_{4m_h^2/s}^1 d\tau \int_\tau^1 dx/x f_{g/P}(x,Q)f_{g/P}(\tau/x,Q)\hat\sigma_{gg\to hh}(\tau s) \ ,
\end{equation}
where the center of mass energy of the collider $s$ is related with $\hat s$ by: $\hat s=\tau s$ and
\begin{align}
\sigma_{gg\to hh}(\hat s)&=\frac{G_F^2\alpha_\text{S}^2(\mu)\hat s}{128(2\pi)^3 9}\sqrt{1-\frac{4m_h^2}{\hat s}}C^2_{\rm eff}(\hat s)\ ,
\\
C_{\rm eff}(\hat s)&=\frac{3m_h^2}{\hat s-m_h^2}\frac{g_h}{2}+\frac{g_{hh}}{2} \ .
\end{align}
$Q$ and $\mu$ are the factorization and renormalization scales that we take as the center of mass energy of the system of two Higgs states: $Q=\mu=\sqrt{\hat s}$. We use the MSTW2008 parton distribution functions for the calculations with LET~\cite{0901.0002}. 

Since $g_h$ and $g_{hh}$ are independent of momentum, using the PDF of gluons we can integrate Eq.~(\ref{eq-xsec-eff}) and obtain an explicit expression for $\sigma_{hh}$ in terms of the parameters of the model as
\begin{equation}\label{eq-xsec-simp}
\sigma_{hh}=\sigma_{hh}^{(a)} g_h^2+\sigma_{hh}^{(b)} g_hg_{hh}+\sigma_{hh}^{(c)} g_{hh}^2 \ ,
\end{equation}
where $\sigma_{hh}^{(j)}$ are integrals of the PDF, independent of the parameters of the NP sector, whereas $g_h$ and $g_{hh}$ contain all the information of the NP, Eqs.~(\ref{eff-gh}) and~(\ref{eff-ghh}).

\begin{figure}[h!]
\centering
\includegraphics[width=0.495\textwidth]{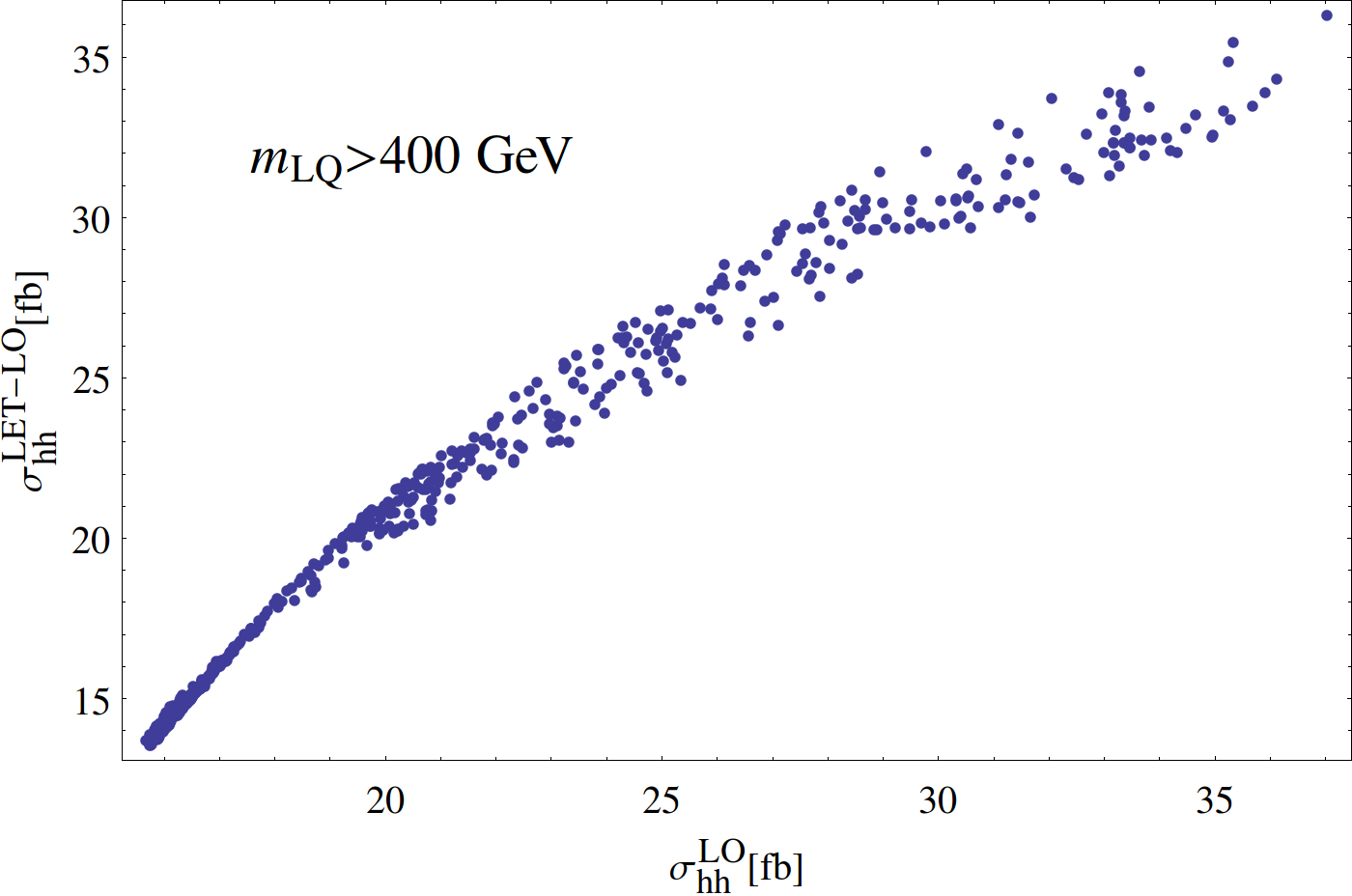}
\includegraphics[width=0.495\textwidth]{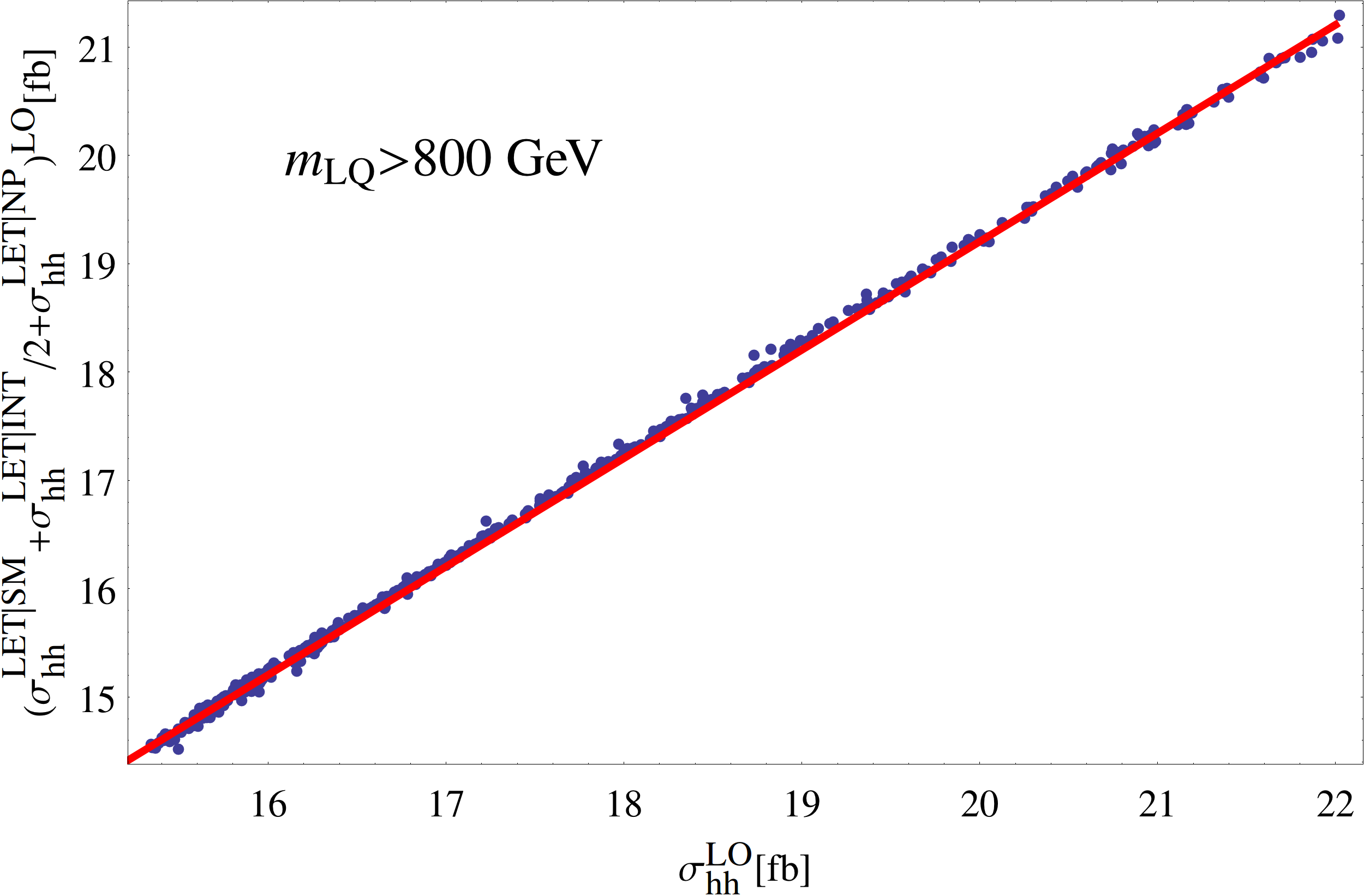}
\caption{On the left we show $\sigma_{hh}$ using LET and full calculation at LO, allowing one of the LQ masses to have a lower bound of 400~GeV, whereas the others are larger than 800~GeV. On the right, considering the LET and all LQ masses larger than 800~GeV, we show $\sigma_{hh}^{\rm LET|SM}+\sigma_{hh}^{\rm LET|INT}/2+\sigma_{hh}^{\rm LET|NP}$, that is: we correct the interference term by a factor 1/2, the red line is the identity, shifted by a constant value to compensate the underestimation of the SM cross section using the LET.}
\label{fig-xsechh-full-vs-LET}
\end{figure}

In Fig.~\ref{fig-xsechh-full-vs-LET} we show the results of the full calculation and the calculation with LET, there are several interesting features to comment. On the left we show the results for a scenario where the mass of one of the LQs is taken as $m_{\rm LQ}^{\rm light}\gtrsim 400$~GeV, whereas the rest of them are larger than 800~GeV. Although the LET correctly predicts the order of magnitude of the cross section, there is a dispersion for large values of $\sigma_{hh}$, that corresponds to $m_{\rm LQ}^{\rm light}$ being near the lowest value. This dispersion can be expected, since for light masses the assumptions for the LET are not valid anymore. There is another correction, that is responsible for the departures of the points from a straight line, that is: for low LQ masses the interference term is sizable, but the LET fails to reproduce it properly. Given that the top mass is not heavy enough compared with other scales of the process, the LET overestimates the interference by a factor $\simeq 2$. On the right we show the case where all the LQ masses are larger than 800~GeV, such that they are heavier than the Higgs mass and the energy flowing through the loop. Unlike the previous case, we have divided by two the interference term, such that the vertical axis corresponds to: $\sigma_{hh}^{\rm LET|SM}+\sigma_{hh}^{\rm LET|INT}/2+\sigma_{hh}^{\rm LET|NP}$. The red line is the identity, with a shift to compensate the top contribution that, in the LET, underestimates the cross section by 20\% approximately.

\begin{figure}[h!]
\centering
\includegraphics[width=0.7\textwidth]{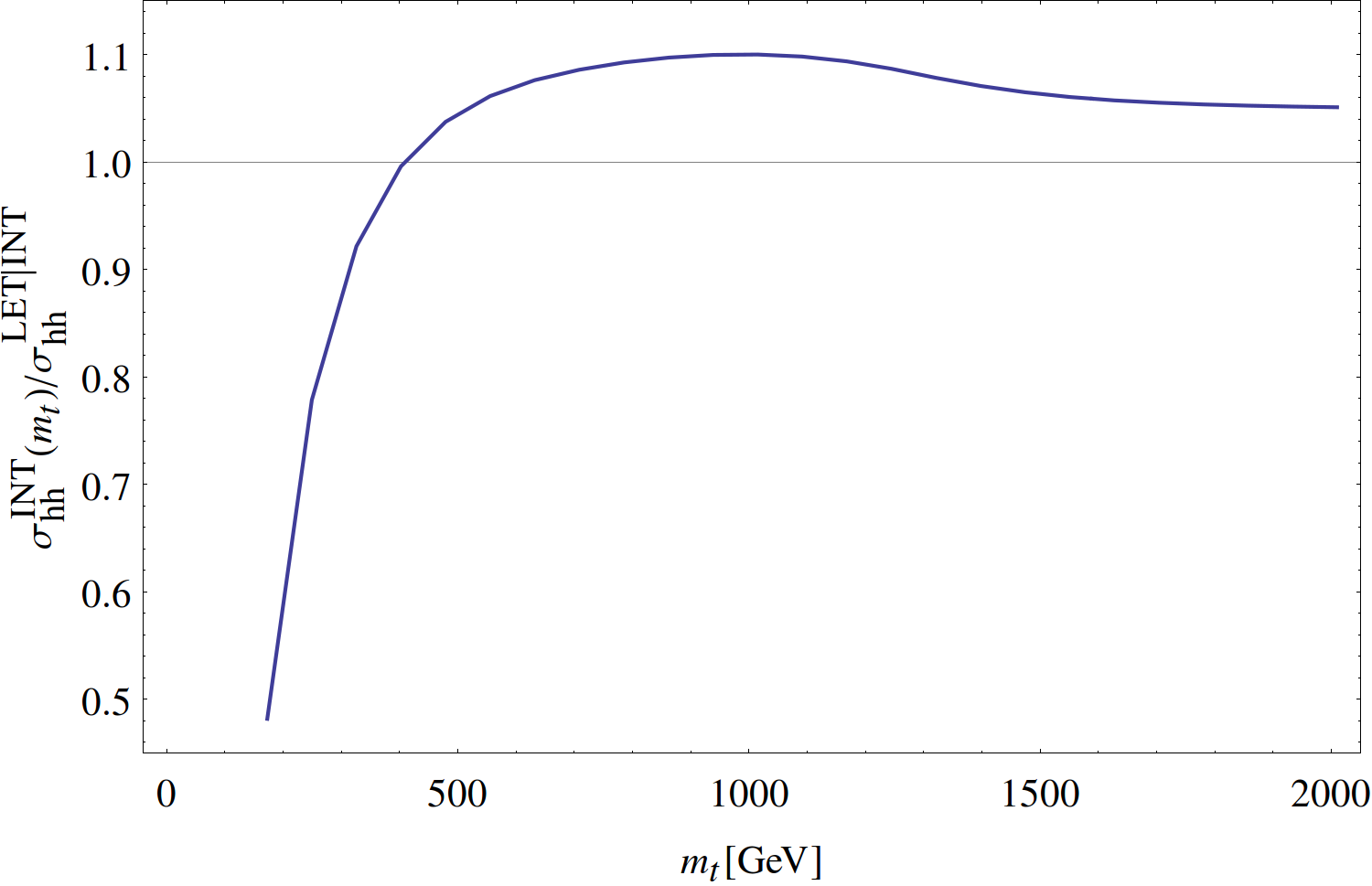}
\caption{$(\sigma_{hh}^{\rm INT}/\sigma_{hh}^{\rm LET|INT})$ to LO as function of the top mass.}
\label{fig-top-limit-xsechh}
\end{figure}

To show that the origin of the factor two in the interference is the heavy top mass approximation, we have checked that, by making the top mass larger, both calculations converge. In Fig.~\ref{fig-top-limit-xsechh} we show the ratio of interference term using the LET and using the full calculation: $\sigma_{hh}^{\rm INT}/\sigma_{hh}^{\rm LET|INT}$, as function of the top mass, for a point of the parameter space with: $m_{\rm LQ}^{\rm light}=1.3$~TeV. This ratio is of order 1/2 for the SM top mass, and converges to 1 for large $m_t$ and LQ masses. In the figure, the curve does not converge to 1 because the LQ masses are finite, but we have checked that as we increase their values it goes to 1.

\begin{figure}[h!]
\centering
\includegraphics[width=0.7\textwidth]{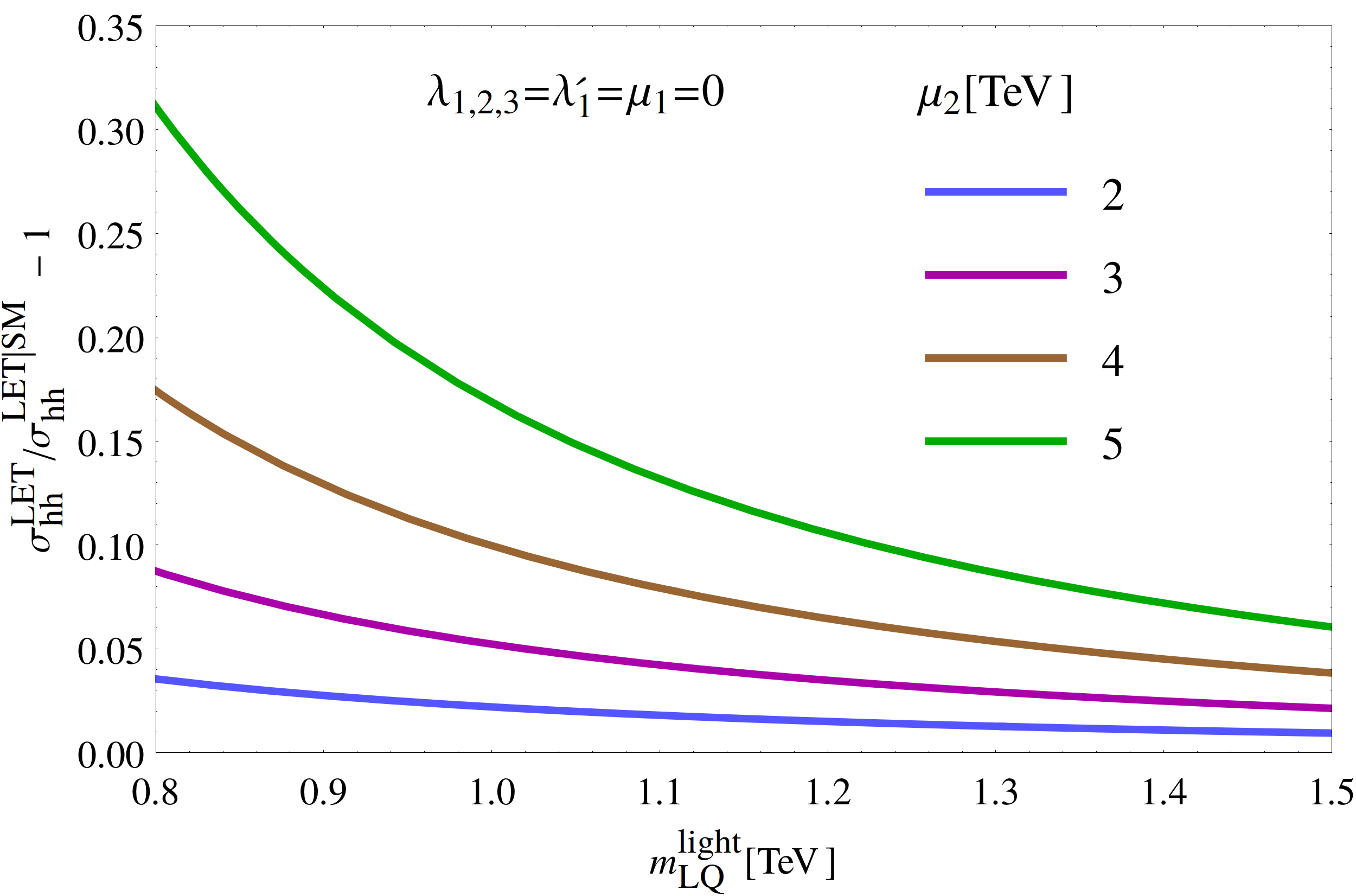}
\caption{$(\sigma_{hh}^{\rm LET}/\sigma_{hh}^{\rm LET|SM})$ - 1 to LO as a function of $m_{\rm LQ}^{\rm light}$, in colors values of $\mu_2$.}
\label{fig-mlight-xsechh}
\end{figure}

In Fig.~\ref{fig-mlight-xsechh} we show only the LQ (SM-subtracted) contribution to the cross section normalized with respect to the SM one, $(\sigma_{hh}^{\rm LET}-\sigma_{hh}^{\rm LET|SM})/\sigma_{hh}^{\rm LET|SM}$, as function of the physical mass of the lightest LQ state. We have taken $\lambda_{1,2,3}=\lambda'_1=\mu_1=0$ and different values for $\mu_2$, as shown by the different curves. We have varied $m_{\bar S_1}$, whereas the other quadratic coefficients are fixed to 2 TeV. For this choice of parameters the mass of the light mass eigenstate is driven by $m_{\bar S_1}$. As expected, for lighter mass and large $\mu_2$, the cross section receives larger corrections.\footnote{We only show masses for which the LET is a reasonable approximation.} Non-vanishing quartic couplings with signs chosen as we have, provide a milder opposite effect than cubic couplings.

\begin{figure}[h!]
\centering
\includegraphics[width=0.85\textwidth]{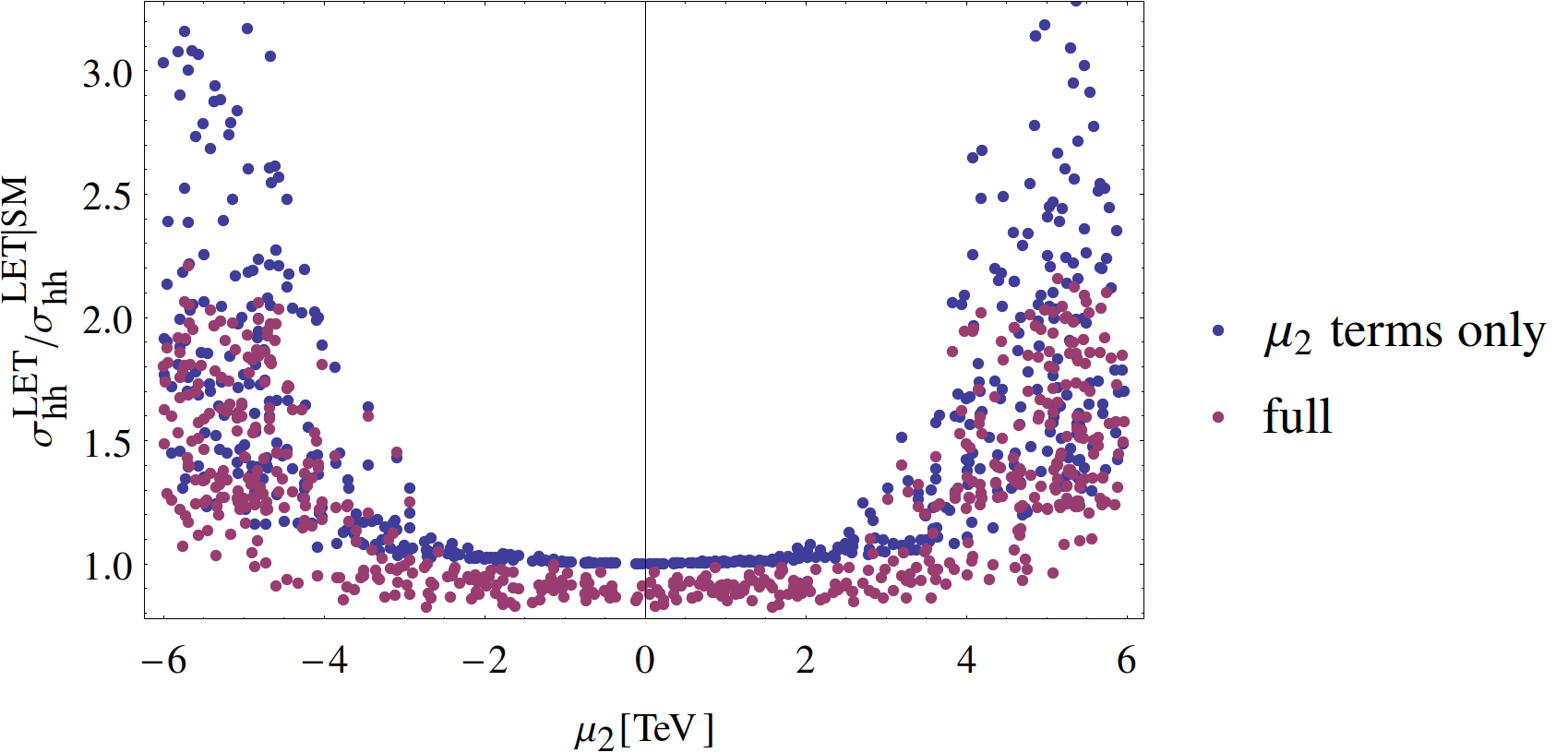}
\caption{$\sigma_{hh}^{\rm LET}/\sigma_{hh}^{\rm LET|SM}$ to LO as a function of $\mu_2$. The purple dots correspond to the full LET computation and the blue ones to the inclusion of $\mu_2$ terms only.}
\label{xsechh-let-mu2approx-lightscenario}
\end{figure}

The Fig.~\ref{xsechh-let-mu2approx-lightscenario} shows the ratio $\sigma_{hh}^{\rm LET}/\sigma_{hh}^{\rm LET|SM}$ vs. $\mu_2$, using the full LET result for the double Higgs production in purple and using the cubic approximations to $g_h$ and $g_{hh}$ in blue, Eqs.~(\ref{ghcubicup}) and~(\ref{ghhcubicup}), respectively, replaced in Eq.~(\ref{eq-xsec-simp}), considering in both cases only contributions from the up-type sector. Note that the approximation does a good job in mimicking the full trend of the LET result, however it tends to overestimate its value. This is expected since in the approximation we are neglecting the quartic contributions which, as mentioned previously, provide an opposite behavior with respect to the cubic ones, decreasing the overall value of the cross section.

\subsection{Higgs couplings}
\label{sec-higgs-couplings}
The LQs interactions with the Higgs can significantly affect the gluon fusion Higgs production and this impact can be quantified through the ratio between the total and the SM cross sections~\footnote{At leading order the LQs effects on the gluon fusion Higgs cross section can be also measured as the ratio between the modified and the SM gluonic partial decay widths of the Higgs ($|\kappa_g|^2 = \Gamma(h\to gg) / \Gamma(h\to gg)^{\rm SM}$) following the relation among the gluon fusion cross section and the partial decay width into gluons~\cite{Djouadi:2005gj}.}                                                                                                                                                                                                                                                                                                                                                                                                                                                                                                                                                                                                                                                                                                                                                                                                                                                                                                                                                                                                                                                                                                                                                                                                                                                                                                                                                                                                                                                                                                                                                                                                                                                                                                                                                                                                                                                                                                                                                                                                                                                                                                                                                                                                                                                                                                                                                                                                                                                                                                                                                                                                                                                                                                                                                                                                                                                                                                                                                                                                                                                                                                                                                                                                                                                                                                                                                                                                                                                                                                                                                                                                                                                                                                                                                                                                                                                      
\begin{equation}
|\kappa_g|^2 = \frac{\sigma(gg \to h)}{\sigma(gg \to h)^{\rm SM}} \ ,
\end{equation}
which is expressed in terms of the top ($A_{1/2}$) and LQs ($A_0$) one-loop functions
\begin{equation}
|\kappa_g|^2 \propto \left|A_{1/2}(\tau_t) + \sum_{k=u,d}\sum_{i=1,2}\frac{v~ \mathscr{C}^{(k)}_{ii}}{2m_{\chi^k_i}^2}A_0(\tau_{\chi^k_i})\right|^2 \ ,
\end{equation}
where $\tau_i = m_h^2 / (4 m_i^2)$ for $i=t,\chi$, and the one-loop functions are normalized as $A_{1/2}(0)=4/3$ and $A_0(0)=1/3$. Defining $\kappa_g=1+\delta \kappa_g$ we separate the LQs contributions
\begin{equation}
\delta\kappa_g = \sum_{k=u,d}\sum_{i=1,2}\frac{v~ \mathscr{C}^{(k)}_{ii}}{2m_{\chi^k_i}^2}\frac{A_0(\tau_{\chi^k_i})}{A_{1/2}(\tau_t)} \simeq 0.12 \sum_{k=u,d}\sum_{i=1,2}\frac{v~ \mathscr{C}^{(k)}_{ii}}{m_{\chi^k_i}^2} \ .
\label{deltakappag}
\end{equation}
The numerical factor in the right hand side results from evaluating all the LQ masses at 800 GeV~\footnote{This is a good approximation even if not all the LQs have a mass of 800 GeV. The extreme case appears in the scenario analyzed in Sec.~\ref{sec-light-scenario} where the mass of the lightest LQ ($\sim$ 400 GeV) produces a correction lower than 1$\%$ to Eq.~(\ref{deltakappag}).}, meaning that the LQs contributions reproduce the ones obtained through the LET discussed in the  previous section, see Eq.~(\ref{ghLET}). The sum is taken over the four LQs physical states. Accordingly, $\mathscr{C}^{(k)}_{ii}$ stand for the trilinear scalar couplings in the physical basis. Following Eqs.~(\ref{eq-V3}) and~(\ref{eq-V4}), we can see that $\mathscr{C}^{(k)}_{ii}$ depends after mixing on the cubic and quartic (once one of the Higgs is evaluated at the vev) scalar couplings in the potential.\footnote{Notice that the cubic couplings in the potential only contribute to the single gluon fusion Higgs production in the physical basis since these interactions involve different LQs in the gauge basis. Conversely, no mixing is necessary for the double Higgs production.}

It is simple to derive an expression that relates the LQ contributions to the single Higgs-digluon coupling, $\delta\kappa_g$, and  the double Higgs cross section, $\sigma_{hh}$, as described in Eq.~(\ref{eq-xsec-simp}) using the LET and considering only a cubic contribution from the up-sector. Noticing that $\delta\kappa_g=(1/8)g^{\rm LQ}_h$, one can write $g_h$ and $g_{hh}$ (note that these include the top quark contribution, see Sec.~\ref{LET}) in terms of $\delta\kappa_g$ as
\begin{eqnarray}
g_h&=&\frac{4}{3}(1+\delta\kappa_g)\label{ghdeltak}\\
g_{hh}&=&-\frac{4}{3}(1-\delta\kappa_g+8\delta\kappa_g^2) \, . \label{ghhdeltak}
\end{eqnarray}
Replacing these expressions in Eq.~(\ref{eq-xsec-simp}) for $\sigma_{hh}$ we get the dependence of the cross section on $\delta\kappa_g$.

Furthermore, the LQs couplings with the Higgs can produce important modifications to the partial decay width of the Higgs into a pair of photons. In this case, these variations are captured by the decay width normalized to its SM value
\begin{equation}
|\kappa_\gamma|^2 = \frac{\Gamma(h\to \gamma\gamma)}{\Gamma(h\to \gamma\gamma)^{\rm SM}} \ ,
\end{equation}
which is determined by the $W$ one-loop function, $A_1$, as well as by $A_{1/2}$ and $A_0$ for the top and LQs contributions, respectively,
\begin{equation}
|\kappa_\gamma|^2 \propto \left|A_{1}(\tau_W)+N_cQ_u^2 A_{1/2}(\tau_t)+N_c\hspace{-5pt} \sum_{k=u,d}\sum_{i=1,2}Q_{\chi^k_i}^2\frac{v~ \mathscr{C}^{(k)}_{ii}}{2m_{\chi^k_i}^2}A_0(\tau_{\chi^k_i})\right|^2 \ ,
\end{equation}
where $N_c$ is the number of colors of the LQs and $Q_\chi$ their electric charges, and $A_1$ is normalized as $A_1(0)=-7$. Analogously to Eq.~(\ref{deltakappag}), we define in this case $\kappa_\gamma=1+\delta \kappa_\gamma$
\begin{align}
\delta\kappa_\gamma & = N_c \sum_{k=u,d}\sum_{i=1,2}Q_{\chi^k_i}^2 \frac{v~ \mathscr{C}^{(k)}_{ii}}{2m_{\chi^k_i}^2}\frac{A_0(\tau_{\chi^k_i})}{A_{1}(\tau_W)+N_cQ_u^2A_{1/2}(\tau_t)}
 \nonumber \\
& \simeq -0.026 N_c\sum_{k=u,d}\sum_{i=1,2}Q_{\chi^k_i}^2 \frac{v~ \mathscr{C}^{(k)}_{ii}}{2m_{\chi^k_i}^2} \ .
\label{deltakappagamma}
\end{align}
The numerical factor in the right hand side results again now from evaluating all the LQ masses at 800 GeV.

From Eqs.~(\ref{deltakappag}) and~(\ref{deltakappagamma}) we see that a combination of large trilinear couplings and small LQ masses produces larger deviations and, consequently, it tends to put these observables under more tension. Both $\kappa_g$ and $\kappa_\gamma$ have been measured at the LHC~\cite{Aad:2019mbh}. We will analyze in the next sections the constraints imposed by the measurements on the parameter space of the model and show the expected correlation between $\kappa_g$ and $\kappa_\gamma$~\cite{Dorsner:2016wpm}.

Finally, another possible important LQ effect might appear in the partial decay width of the Higgs into a $Z$ and a photon, $\Gamma(h\to Z\gamma)$. In this case, however, we expect a contribution of the same order as in $\Gamma(h\to \gamma\gamma)$~\cite{Dorsner:2016wpm} with smaller experimental restrictions~\cite{Aad:2020plj}. As we show in Sec.~\ref{higgscoouplings}, $\Gamma(h\to \gamma\gamma)$ introduces no more than slight constraints on the parameter space in regions where the double Higgs production is enhanced, therefore, $\Gamma(h\to Z\gamma)$ ends up having a negligible impact on our analysis.

\section{Constraints}\label{constraints}

In this section we analyze the restrictions imposed on the model by different observables. In particular, we concentrate on the constraints arising from the oblique corrections to electroweak precision tests, LQs contributions to the $Zb\bar{b}$ coupling, flavor changing transitions and LQ direct searches. We also study the requirements for baryon and lepton number conservation and define the scenarios of interest for the phenomenological analysis.

\subsection{Oblique corrections}\label{sec-oblique}
The LQs give contributions to the self-energies of the EW gauge bosons at one-loop level~\cite{0603188,0904.1625,Dorsner:2016wpm}. The leading constraints to these quantities are captured by the $S$ and $T$ parameters~\cite{Peskin-Takeuchi,Altarelli-Barbieri}: $S=0.05\pm 0.11$ and $T=0.09\pm 0.13$.
 
There are two sources for the $T$-parameter: the splitting of the doublet induced by $\lambda_1'$ and the mixing induced by $\mu_{1,2}$. Defining
\begin{equation}
f_T(x,y)= \frac{xy}{x-y}\log\frac{x}{y} \, ,
\end{equation}
and using Eqs.~(\ref{eq-rot1}) and (\ref{eq-rot2}), we obtain
\begin{align}
\Delta T=\frac{N_c}{16\pi c_W^2s_W^2m_Z^2}\sum_{i,j=1}^2\left[
2f_T(m_{\chi^u_i}^2,m_{\chi^d_j}^2)|U_{u,1i}^*U_{d,1j}|^2
-\sum_{x=u,d}f_T(m_{\chi^x_i}^2,m_{\chi^x_j}^2)|U_{x,1i}^*U_{x,1j}|^2\right] \ .
\end{align}
Notice that for $m_\chi\sim{\cal O}(0.5-2)\ {\rm TeV}$ and splittings between the components of a doublet of ${\cal O}(1)$, $T$ can take values $\gg 1$, thus $T$ is expected to give interesting constraints on the parameter space of the model.

In order to understand how $\lambda_1'$ and $\mu_{1,2}$ generate contributions to $T$, one can consider the limit $\mu_{1,2} v\ll m_{\tilde R_2}^2$
\begin{align}
\Delta T\simeq&\frac{N_c}{16\pi c_W^2s_W^2m_Z^2}
\left[\frac{2 v^4 \lambda_1'^2}{6 m_{\tilde R_2}^2 + 3 v^2 \lambda_1}
+ \frac{[(m_{\tilde R_2}^2 - m_{\bar S_1}^2) \mu_1^2 + (-m_{\tilde R_2}^2 + m_{S_1}^2) \mu_2^2]^2 v^4}{12 m_{\tilde R_2}^2 (m_{\tilde R_2}^2 - m_{S_1}^2)^2 (m_{\tilde R_2}^2 - m_{\bar S_1}^2)^2}
\right.\nonumber\\
&\left.+ \frac{2 v^4 \lambda_1'}{3 (m_{\tilde R_2}^2 + v^2 \lambda_1/2)} \left(\frac{-\mu_1^2}{2 m_{\tilde R_2}^2 - 2 m_{S_1}^2 + v^2 (\lambda_1 - \lambda_2)} + \frac{\mu_2^2}{2 m_{\tilde R_2}^2 - 2 m_{\bar S_1}^2 + v^2 (\lambda_1 - \lambda_3)}\right)
\right] \ ,
\label{eq-aproxT}
\end{align}
where we show the first corrections generated by $\lambda_1'$ and $\mu_{1,2}$ in the first line, and in the second line a mixed correction. We have checked by comparison with the full result at a numerical level that, in the proper region of parameter space, this approximation gives a good estimate of $T$.

Similar expressions can be derived for the $S$-parameter. Defining
\begin{equation}
f_S(x,y)= -1+\frac{x\log x-y\log y}{x-y} \, ,
\end{equation}
we obtain
\begin{align}
\Delta S=-\frac{Y}{2\pi}\sum_{i,j=1}^2
\left[f_S(m_{\chi^u_i}^2,m_{\chi^u_j}^2)|U_{u,1i}^*U_{u,1j}|^2
-f_S(m_{\chi^d_i}^2,m_{\chi^d_j}^2)|U_{d,1i}^*U_{d,1j}|^2\right] \ .
\end{align}
We do not expect large corrections to $S$. Considering an example, for a single doublet: $\Delta S=-(Y/2\pi) \log(m_{\chi^u}^2/m_{\chi^d}^2)$, that for $Y=1/6$ and $(m_{\chi^u}^2/m_{\chi^d}^2)\in(0.1,10)$ gives $\Delta S\in(-0.06,+0.06)$. For our model and region of parameter space $S$ does not lead to important constraints.

\subsection{$Z$-couplings}
LQs also give corrections at one-loop level on vertex interactions of SM fermions and EW gauge bosons~\cite{9411392,1412.8480}. Precision measurements at LEP give very stringent constraints on the modifications of the $Z$-couplings~\cite{0612034}. Focussing on $Zb\bar b$, Ref.~\cite{Dorsner:2016wpm} has shown that for $m_{\rm LQ}\simeq 0.5\ {\rm TeV}$, the couplings to fermions are required to be $\lesssim 10$. Since in our approach those couplings will be taken small, the constraints from $Z$-couplings can be easily avoided.

\subsection{Direct searches}\label{directsearches}
Bounds from direct searches of LQs at colliders depend on the final state. For pair production, that is dominated by QCD, these bounds are usually presented in terms of the BRs.

Final states with top quarks and $e$ or $\mu$ have stronger bounds, with $m_{\rm LQ}\gtrsim 1.6\ {\rm TeV}$ for BRs to light leptons $\simeq 1$~\cite{2010.02098}. LQs dominantly coupled to light quarks and leptons are also excluded up to $m_{\rm LQ}\gtrsim 1.6\ {\rm TeV}$~\cite{1605.06035,1906.08983}. For decays to quarks and leptons of the third generation: $m_{\rm LQ}\gtrsim 800-900\ {\rm GeV}$~\cite{1902.08103}. For the decay to a pair of down type quarks, \cite{1710.07171} excludes masses below 410 GeV and 610 GeV for decays to $qq$ and $bq$, respectively.

The above searches assume that the LQs decay promptly and no longer apply if they are long lived. In that case, bounds arising from searches of long lived squarks must be considered instead. Different final states involving decays into down-type quarks and a lepton~\cite{Aad:2020srt,Sirunyan:2020cao} or two down-type quarks~\cite{1902.01636,Sirunyan:2021kty,Sirunyan:2020cao} have been explored. All the exclusion limits reported in these searches restrict the long-lived particle mass to be well above 1 TeV. For example, in~\cite{Sirunyan:2020cao} top squark masses up to 1.6 TeV are excluded for mean proper decay lengths between 3 and 300 mm.

\subsection{Baryon and Lepton numbers}\label{sec-BL}
The interactions of LQs can violate baryon and lepton number, inducing processes forbidden in the SM, as proton decay and neutron-antineutron oscillations. In fact, if one includes all the renormalizable interactions conserving the gauge symmetries of the SM and the Poincar\'e symmetry, then $B$ and $L$ are violated. Below we discuss different cases where these symmetries can be conserved, as well as situations in which they are violated, but proton decay can be sufficiently suppressed. In the rest of the paper we will concentrate in the first and second cases of the $B$ and $L$ conserving scenarios.

\subsubsection{$B$ and $L$ conservation}\label{sec-BLconservation}
Taking different combinations of vanishing couplings in Eq.~(\ref{eq-ferm-int}), the LQs can be assigned $B$ and $L$ charges such that these numbers are conserved.

\begin{enumerate}
\item Taking $y_{2\ell}=z_{S_1q_L}=z_{S_1q_R}=\mu_1=0$, one can asign:
$B(\tilde R_2)=-B(\bar S_1)=2B(S_1)=-2/3$, $L(\tilde R_2)=L(\bar S_1)=0$ and $L(S_1)=-1$. Notice that in this case $\tilde R_2$ and $\bar S_1$ do not interact with leptons at the level of dimension four operators. For phenomenology this is a very interesting situation, since it allows one of the cubic interactions, whereas $\tilde R_2$ and $\bar S_1$ decay to dijets and $S_1$ decays to $q\ell$. These assignments and a lighter LQ, with a mass $\sim 400-600\ {\rm GeV}$, define our {\it scenario 1}, dubbed ``Scenario with a light LQ", and it will be analyzed in Sec.~\ref{sec-light-scenario}.
\item Taking vanishing interactions with fermions, except for $y_{2\ell}$, one can assign:
$B(\tilde R_2)=-B(\bar S_1)=-B(S_1)=1/3$ and $L(\tilde R_2)=-L(\bar S_1)=-L(S_1)=-1$. In this case $S_1$ and $\bar S_1$ decay to $q\ell$ after mixing. This is also a very interesting situation, since it allows both cubic interactions, as well as decay of all the LQs. These assignments and LQs with masses $\gtrsim 800\ {\rm GeV}$, define our {\it scenario 2}, dubbed ``Scenario with heavy LQs", and it will be analyzed in Sec.~\ref{sec-heavy-scenario}.
\item Taking $z_{\bar S_1 d}=z_{S_1q_L}=z_{S_1q_R}=\mu_1=0$, one can assign:
$B(\tilde R_2)=-B(\bar S_1)=-B(S_1)=1/3$ and $L(\tilde R_2)=-L(\bar S_1)=L(S_1)=-1$. Diquark interactions are not present and $\bar S_1$ decays after mixing with the up-component of $\tilde R_2$ to $q\ell$. Regarding the LQ-Higgs interactions at leading order, this scenario is already included in the ``Scenario with heavy LQs". However, it is disfavored from the phenomenological perspective since it is constrained by the same bounds on the LQ masses arising from the direct searches and the condition $\mu_1=0$ implies a smaller contribution to the double Higgs production.
\end{enumerate}

\subsubsection{$B$ and $L$ violation} 
In the general case dimension six operators as $ddu\bar \nu$ can be induced at 1-loop level, as well as dimension nine operators as $ddu\bar due$ at tree level~\cite{1204.0674}. They can produce $n\to\pi^0\nu,\pi^+e$, $n\bar n$ oscillations and most dangerous: $p\to\pi^+\pi^+e^-$. Proton decay is proportional to $|y_{2\ell}z_{\bar S_1 d}|^2/m_{\rm LQ}^4$, thus it can be suppressed by taking $y_{2\ell}=0$ for $d$ and $s$, leaving only coupling with $b$, or taking $y_{2\ell}$ small enough to satisfy the constraints.

\subsection{Flavor changing transitions}
Since the LQs interact with the SM fermions, they can induce flavor transitions and CP violation. At tree level they can contribute to processes involving quarks and leptons, as semileptonic and leptonic decays of mesons, decay of heavy leptons as the $\tau$ and transitions in nuclei. At loop level they can also contribute to processes as meson mixing, dipole moments and rare decays. The absence of conclusive evidence of physics BSM in the flavor sector strongly constrains the flavor structure and the size of the couplings to fermions. Below we give estimates for the most constraining processes, that usually involve light fermions, and we refer to~\cite{Dorsner:2016wpm} for an extended list of processes as well as the details of the calculations.

Semileptonic interactions can be generated by tree level exchange of LQs, their coefficients being estimated to be of order $C_{4f}\sim y^2/m_{\rm LQ}^2$, with $y$ a generic coupling to fermions. Leptonic decays of pseudoscalar mesons are some of the most sensitive processes receiving contributions from these interactions. The ratio $R^\pi_{\mu/e}=\Gamma(\pi\to e\nu)/\Gamma(\pi\to \mu\nu)$ can be predicted with good precision in the SM due to cancellation of uncertainties, leading after comparison with the experimental result to the estimates~\cite{1009.3886}: $|{\rm Re}(y_{S_1e,11}^*y_{S_1\ell,11})|\lesssim 10^{-7}(m_{\rm LQ}/{\rm TeV})^2$ and $|{\rm Re}(y_{S_1e,12}^*y_{S_1\ell,12})|\lesssim 2\times 10^{-3}(m_{\rm LQ}/{\rm TeV})^2$.

Four-fermion operators involving only quarks are generated at one-loop level, with Wilson coeficients that roughly can be estimated as: $C_{4f}\sim y^4/(4\pi m_{\rm LQ})^2$~\cite{9309310}. To get an idea on the size of the allowed couplings, in the absence of flavor symmetries, one can consider one of the most stringent constraints that arises from $\epsilon_K$ in the Kaon system, where for complex couplings $C_{4f}\lesssim 7\times 10^{-9} \; {\rm TeV}^{-2}$~\cite{1002.0900}. Assuming that $d$ and $s$ quarks have couplings of similar size, this bound leads to $y\lesssim 6\times 10^{-3}$, when $m_{\rm LQ}\simeq 500\ {\rm GeV}$.\footnote{One could think that diquark interactions could contribute to $C_{4f}$ at tree level, however antisymmetry of $z$-couplings cancel these effects~\cite{Dorsner:2016wpm}.} These bounds are relaxed by one order of magnitude if the couplings are real, and at least by one order of magnitude for mesons involving heavy quarks.

The electromagnetic dipole moment (EDM) of the electron has a very tight constraint. LQs with couplings to Left and Right fermions can induce corrections to the EDM at one-loop level~\cite{1304.6119}, leading to strong bounds on the imaginary component of the couplings to electrons, as can be the case of $S_1$. Following~\cite{Dorsner:2016wpm} leads to: ${\rm Im}[y_{S_1e,31}(V^ty_{S_1\ell})_{31}^*]\lesssim 10^{-12}$, where the contribution is dominated by the top quark in the loop due to the requirement of a chiral flip. Neglecting Cabibbo suppressed terms and assuming anarchic complex couplings of the same order, one obtains: $y_{S_1e/\ell,31}\lesssim 10^{-6}$. However these bounds are relaxed by many orders of magnitude for quarks and leptons of second and third generations. Notice that the previous bound requires the presence of two different Yukawa interactions, thus, in the absence of two couplings, the contribution is instead suppressed by the mass of the lepton, eliminating the tension for the case of the electron due to its small mass.

Flavor violating leptonic decays $\ell\to\ell'\gamma$ also have strong bounds. LQs with Left and Right couplings give contributions with chiral enhancement dominated by the top, as well as logarithmic enhancement, whereas those with one coupling are proportional to the lepton masses instead. Following Refs.~\cite{1304.6119,1009.3886} and using bounds for the branching ratios from the PDG~\cite{Zyla:2020zbs}, we obtain the following estimates: (a) for $\mu\to e\gamma$: $|y_{S_1e,31}^*y_{S_1\ell,32}|,|y_{S_1e,32}^*y_{S_1\ell,31}|\lesssim 10^{-7}(m_{\rm LQ}/{\rm TeV})^2$ and $|y_{2\ell,k1}^*y_{2\ell,k2}|\lesssim 0.5\times 10^{-3}(m_{\rm LQ}/{\rm TeV})^2$, (b) for $\tau\to \ell\gamma$ the bounds are of the same order for $\ell=e,\mu$: $|y_{S_1e,3\ell}^*y_{S_1\ell,33}|,|y_{S_1e,33}^*y_{S_1\ell,3\ell}|\lesssim 3\times 10^{-4}(m_{\rm LQ}/{\rm TeV})^2$ and $|y_{2\ell,k\ell}^*y_{2\ell,k3}|\lesssim 0.1(m_{\rm LQ}/{\rm TeV})^2$.
 
LQs also contribute to decays $\ell\to\ell'\ell''\ell'''$, with $\log(m_q^2/m_{\rm LQ}^2)$ enhancement from gamma-penguin diagrams. Following Ref.~\cite{9911539} we have estimated the bound on the couplings for $\mu\to 3e$, that has the strongest constraint, obtaining: $|y_{S_1e/\ell,k1}^*y_{S_1e/\ell,k2}|\lesssim 3\times 10^{-3}(m_{\rm LQ}/{\rm TeV})^2$ and $|y_{2\ell,k1}^*y_{2\ell,k2}|\lesssim \times 10^{-2}(m_{\rm LQ}/{\rm TeV})^2$.

$\mu-e$ conversion in nuclei receives contributions from exchange of LQs at tree and loop levels. Following Refs.~\cite{1009.3886,Dorsner:2016wpm}, the strongest constraints are for couplings with quarks of first generation: $|y_{11}y_{12}^*|\lesssim 0.7\times 10^{-7}(m_{\rm LQ}/{\rm TeV})^2$.

LQs can also give contributions to the anomalous magnetic moments at one-loop level. Following~\cite{Dorsner:2016wpm}, an explanation of the anomalous magnetic moment of the muon with NP at TeV can be achieved in the presence of chiral enhancement, as in the case of $S_1$ with: ${\rm Re}(y_{S_1e,32}^*y_{S_1\ell,32})\sim 3\times 10^{-3}(m_{\rm LQ}/{\rm TeV})^2$. For $\tilde R_2$ there is no chiral enhancement due to the presence of just one coupling, leading to much larger couplings.

At first order the effect of LQs on the Higgs physics is independent of these couplings, thus we will assume that they are small enough to avoid issues with flavor. In the limit of vanishing couplings the LQs are stable, by demanding their decay length to be smaller than $\ell_{\rm min}$ requires $y,z\gtrsim 10^{-6}\sqrt{({\rm TeV}/m_{\rm LQ})^2(0.1{\rm mm}/\ell_{\rm min})(E/6.5{\rm TeV})}$, where $E$ is the energy of the LQ that has been taken of order 6.5 TeV, that is the full available energy at LHC13 in pair production. Thus we will assume that at least one of the couplings that allows the decay of the LQs is on the proper window to satisfy flavor bounds and simultaneously allow a decay length smaller than 0.1mm. In this manner, the LQ masses considered in scenarios 1 and 2 are allowed by the direct searches at colliders since we avoid the more stringent constraints arising from long-lived particles searches (see Sec.~\ref{directsearches}).

In Secs.~\ref{sec-light-scenario} and~\ref{sec-heavy-scenario} we will discuss some specific scenarios, showing configurations of couplings with fermions that are compatible with these conditions.

\section{Scenario with a light LQ}\label{sec-light-scenario}
In this section we show the predictions for several interesting quantities in the scenario 1 of Sec.~\ref{sec-BLconservation}. This scenario allows a light LQ, with a mass $\sim 400-600$~GeV, decaying to dijets. Since its decay to bottom quarks must be suppressed, the simplest choice is to take $z_{\bar S_1d,i3}=0$ in Eq.~(\ref{eq-ferm-int}), with $z_{\bar S_1d,12}$ driving the decays of $\bar S_1$ and $\tilde R_2$. The up-type components are mixed by $\mu_2$, whereas the down-type LQs do not mix, since $\mu_1=0$. $S_1$ interacts with the other scalars through quartic couplings or exchange of other particles. The simplest choice is to assume interactions of $S_1$ with fermions of the third generation only.

We have scanned over the parameter space of the model, in the following range: $m_{\tilde R_2}\in(1.9,2.5)\ {\rm TeV}$, $m_{S_1}\in(0.8,3.5)\ {\rm TeV}$, $m_{\bar S_1}\in(0.4,0.8)\ {\rm TeV}$, $\lambda_{1,3}\in(0,4)$, $\lambda_2\in(3,5)$, $\lambda_1'\in(0,1.5)$ and $|\mu_2|\leq 6\ {\rm TeV}$.

We have kept only points with the lightest up-type LQ being heavier than 400 GeV and the rest of the LQs being heavier than 800~GeV. For the points that we show the mass of lightest state takes values $400-837$~GeV, with most of the points corresponding to masses lighter than 600 GeV, as can be seen in Figs.~\ref{fig-mlight-xsechh-lightscenario} and \ref{fig-dkg-xsechh-lightscenario}. We find that the heavy up-type state has a mass $\sim 1.9-2.5$~TeV, the light down-type state in general is identified with $S_1$, except in some cases for masses above 2 TeV, and it has mass $\sim 0.86-2.52$~TeV, whereas the heavy down-type LQ has mass $\sim 1.9-3.5$~TeV, and in general corresponds to $R_2^d$. The mixing angle in the up-sector is smaller than 0.26.

\subsection{Double Higgs production}

We show in this section the most interesting results from the numerical scan for the scenario with a light LQ using the physical notion from the LET limit where possible.

\begin{figure}[h!]
\centering
\includegraphics[width=0.51\textwidth]{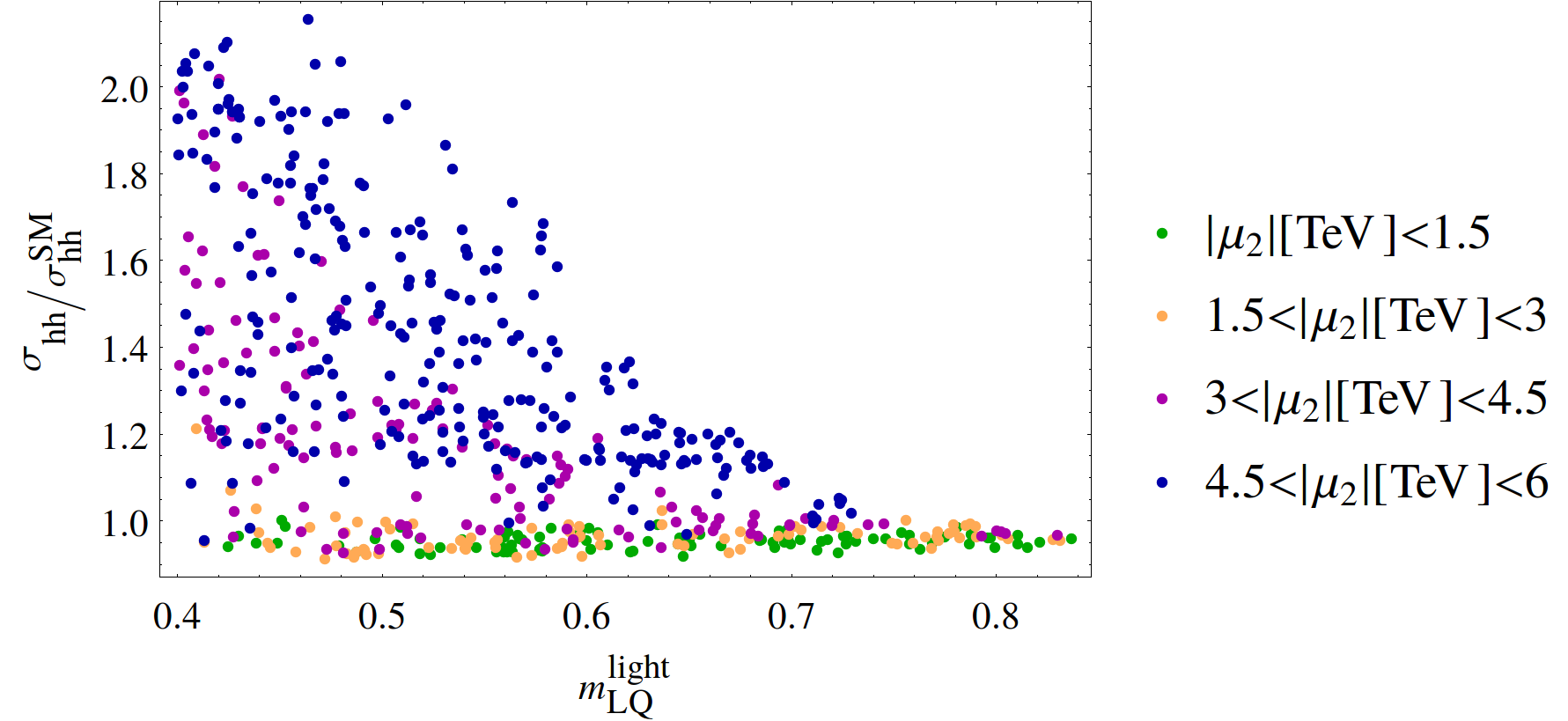}
\includegraphics[width=0.47\textwidth]{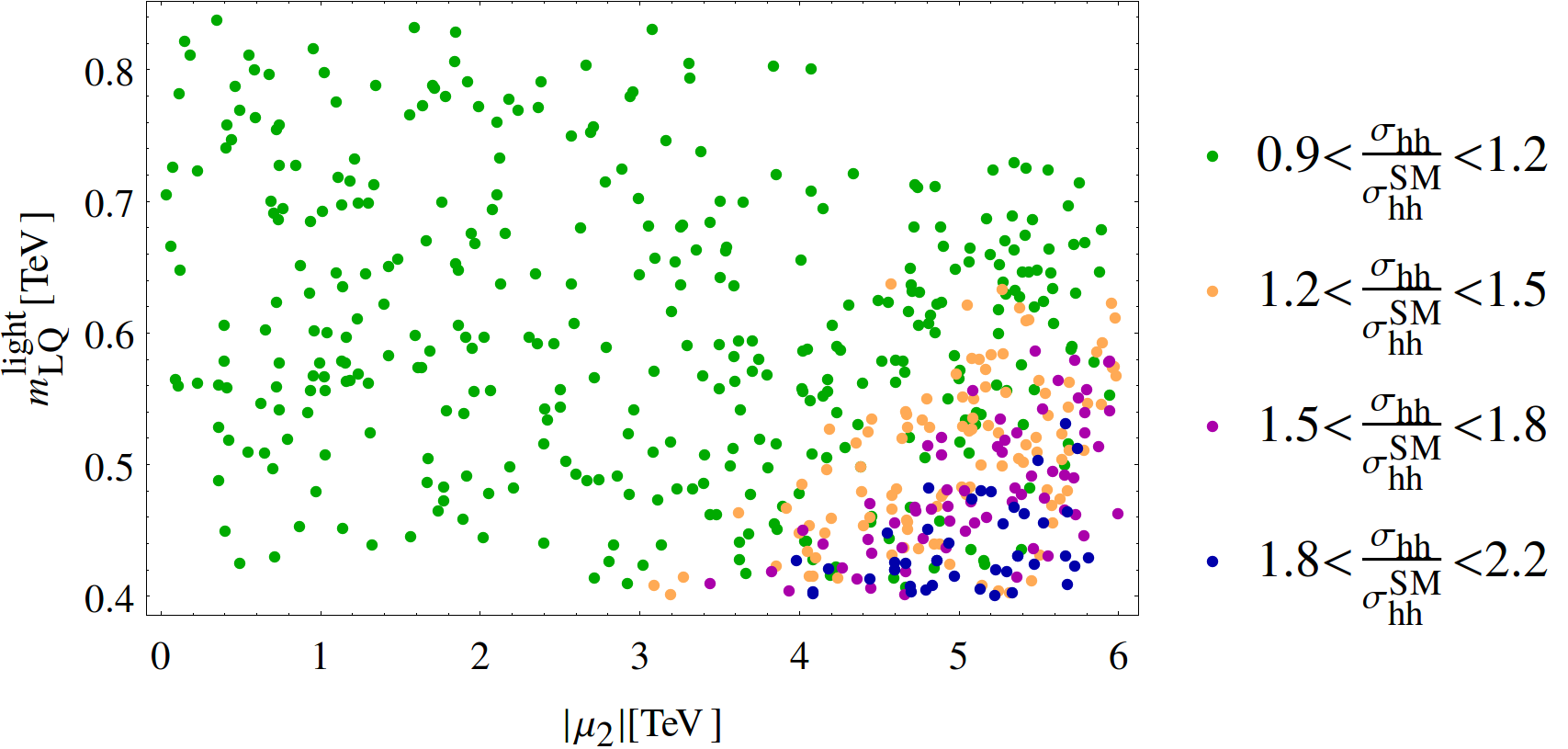}
\caption{Left: LO $\sigma_{hh}$ normalized with respect to the SM one, as function of the lightest LQ mass $m_{\rm LQ}^{\rm light}$ in TeV, in colors the values of $|\mu_2|$. Right: LO $\sigma_{hh}$ normalized with respect to the SM one, as function of $\mu_2$ and $m_{\rm LQ}^{\rm light}$, in colors the values of $\sigma_{hh}/\sigma^{\rm SM}_{hh}$.}
\label{fig-mlight-xsechh-lightscenario}
\end{figure}

In Fig.~\ref{fig-mlight-xsechh-lightscenario} on the left we show the ratio between the total cross section and the SM cross section $\sigma_{hh}/\sigma^{\rm SM}_{hh}$ against the lightest LQ mass, $m_{\rm LQ}^{\rm light}$, in TeV. It is clear from the figure that, as the lightest LQ mass decreases, the possibility of enlarging the total cross section increases, reaching values of order $2\sigma^{\rm SM}_{hh}$ for LQ masses $m_{\rm LQ}^{\rm light}< 0.5$ TeV. However, lighter LQ masses do not necessarily imply larger cross sections. Also notice that when $m_{\rm LQ}^{\rm light}>0.7$ TeV the cross sections are less than a 20\% larger than the SM cross section. We have colored the points in order of increasing value of $|\mu_2|$ from green to blue. Notice that the low values of $|\mu_2|$ (green points) imply smaller cross sections whereas the points with the largest cross sections are associated with the largest values of $|\mu_2|$. In Fig.~\ref{fig-mlight-xsechh-lightscenario} on the right we plot $m_{\rm LQ}^{\rm light}$ in TeV vs. $|\mu_2|$ in TeV, coloring the points in order of increasing value of $\sigma_{hh}/\sigma^{\rm SM}_{hh}$ from green to blue. This figure directly shows that the largest cross sections are accomplished in the regions of small $m_{\rm LQ}^{\rm light}$  and large $|\mu_2|$.

\begin{figure}[h!]
\centering
\includegraphics[width=0.65\textwidth]{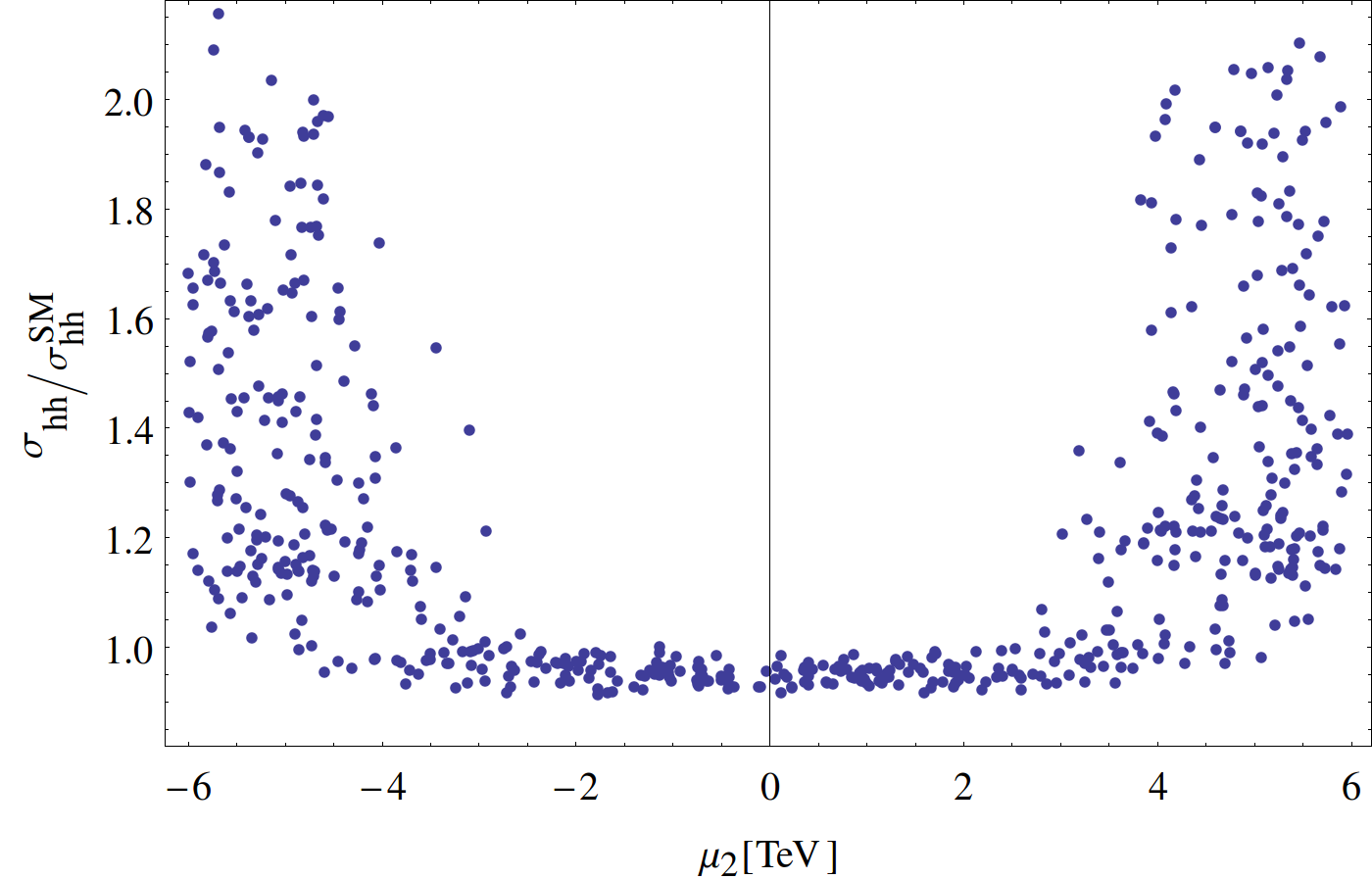}
\caption{LO $\sigma_{hh}$ normalized with respect to the SM one, as function of $\mu_2$.}
\label{fig-mu2-xsechh-lightscenario}
\end{figure}

In Fig.~\ref{fig-mu2-xsechh-lightscenario} we show  $\sigma_{hh}/\sigma^{\rm SM}_{hh}$ vs. $\mu_2$ in TeV. Notice that the increments in the total cross section can be accomplished for values of $|\mu_2|> 3$ TeV, roughly independently of the sign of $\mu_2$. As larger values of $\mu_2$ are considered, the increment in the total cross sections can be sharp for some of the points, following a quartic dependence on $\mu_2$, as we saw as well for the LET case.

\begin{figure}[h!]
\centering
\includegraphics[width=1.\textwidth]{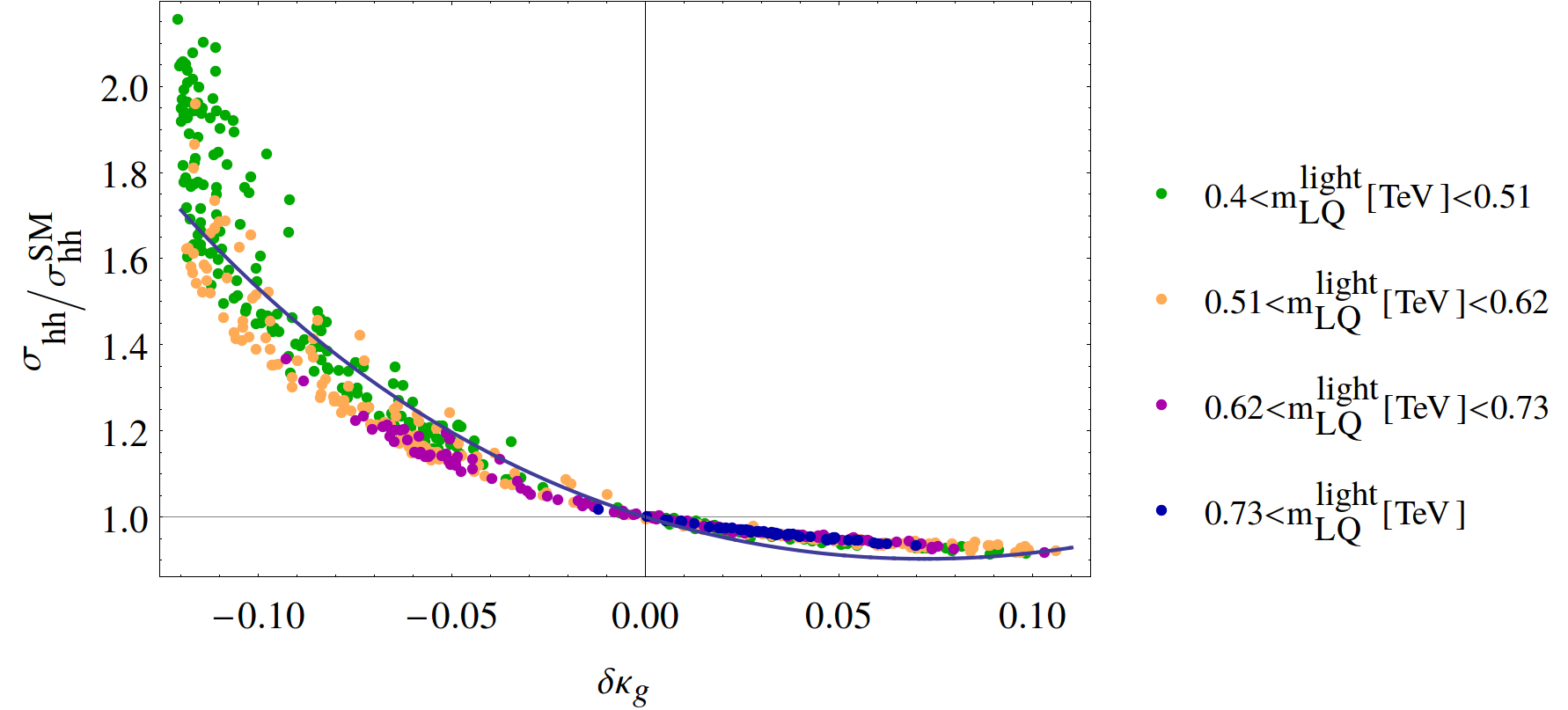}
\caption{LO $\sigma_{hh}$ normalized with respect to the SM one, as function of $\delta\kappa_g$, in color values of the lightest LQ mass $m_{\rm LQ}^{\rm light}$. The blue line corresponds to the LET result.}
\label{fig-dkg-xsechh-lightscenario}
\end{figure}

In Fig.~\ref{fig-dkg-xsechh-lightscenario} we show  $\sigma_{hh}/\sigma^{\rm SM}_{hh}$  vs. $\delta \kappa_g$, coding the colored points in order of increasing value of $m_{\rm LQ}^{\rm light}$ from green to blue, including the LET result depicted by the blue line for $\sigma_{hh}^{\rm LET}(\delta\kappa_g)$ in the cubic dominated case as given in Eq.~(\ref{eq-xsec-simp}) with the replacements of Eqs.~(\ref{ghdeltak}) and~(\ref{ghhdeltak}). We can clearly see that the LET describes nicely the main trend of the figure, in particular for most of the points with  $m_{\rm LQ}^{\rm light}> 0.51$ TeV, for which the LET works well and in which it has been proven the strong correlation between the double Higgs cross section and the modifications to the Higgs-gluon coupling. The deviations from this behavior can be seen for the points that spread vertically for a fixed value of $\delta \kappa_g$, and are represented by the green points which form a sort of a triangular region above the curve. This difference from the LET behavior happens due to the lightest LQ mass decreasing, entering a regime in which the LET breaks down and is unable to capture the full picture. Note also that the vertical break at $\delta \kappa_g \approx -0.12$ corresponds to the constraint on single Higgs production as depicted in Fig.~\ref{fig-dkgamma-dkg-lightscenario} and that for  $m_{\rm LQ}^{\rm light} > 0.7$ TeV, $\sigma^{\rm SM}_{hh}\lesssim \sigma_{hh}\lesssim 1.2 \sigma^{\rm SM}_{hh} $ which agrees with what we find using the LET as shown in Fig.~\ref{fig-mlight-xsechh} once the quartic contributions are accounted for, which tend to contribute oppositely to the cubic, lowering the cross section.

We can estimate at what luminosities the $\sqrt{s}=14$ TeV LHC would be providing the first hints of double Higgs production for the largest cross sections predicted in this scenario with a light LQ. For that purpose, we use the projections by ATLAS~\cite{TheATLAScollaboration:2014scd,TheATLAScollaboration:2015rzr,ATLAS:2016qjh} and CMS~\cite{CMS:2017cwx} that consider different final states but under the best case scenarios seem to be able to constrain $\sigma_{hh}\gtrsim 1.5$ $\sigma^{\rm SM}_{hh}$ at 95$\%$ Confidence Level and at a luminosity of $\mathcal{L}\sim3$ ab$^{-1}$. Given that the largest cross sections we can accomplish within this scenario are $\sigma_{hh}\sim 2.3$ $\sigma^{\rm SM}_{hh}$, a relation we expect not to change sizeably after taking higher order corrections, and that the final states are the same as in the SM case, we estimate that for a luminosity $\mathcal{L}\sim2$ ab$^{-1}$, the first hints (or the exclusion at 95$\%$ Confidence Level) should be able to be seen at the LHC.

\subsection{Higgs couplings}\label{higgscoouplings}
We study next the correlation between $\kappa_g$ and $\kappa_\gamma$. We also consider the dependence of $\kappa_g$ on the cubic coupling of the light up-type state which dominates the LQs contributions within this scenario.

\begin{figure}[h!]
\centering
\includegraphics[width=0.85\textwidth]{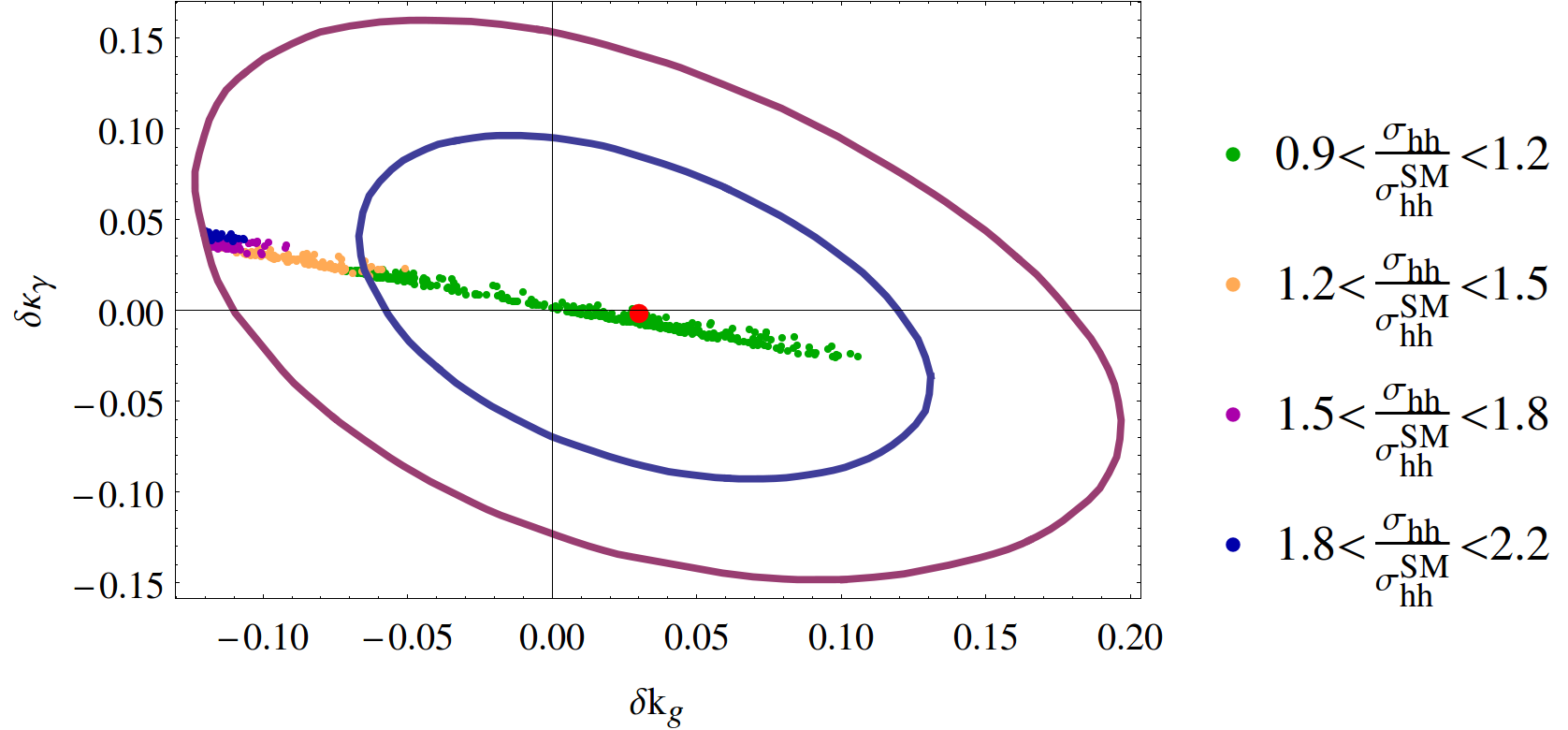}
\caption{Values of $\delta \kappa_\gamma$ and $\delta \kappa_g$ for the points of the scan, in colors the values of $\sigma_{hh}/\sigma^{\rm SM}_{hh}$. The blue and violet lines correspond to the 1$\sigma$ and 2$\sigma$ experimental contours from~\cite{Aad:2019mbh}. The red dot is the central value of the measurement.}
\label{fig-dkgamma-dkg-lightscenario}
\end{figure}

We show in Fig.~\ref{fig-dkgamma-dkg-lightscenario} $\delta \kappa_\gamma$ vs. $\delta \kappa_g$ and the corresponding 1$\sigma$ and 2$\sigma$ experimental contours~\cite{Aad:2019mbh}. A first observation is that $\kappa_g$ enforces the stronger constraints, whereas $\kappa_\gamma$ plays a weaker role in imposing restrictions. In order to evaluate the double Higgs cross section, we have allowed deviations up to the 2$\sigma$ level in this plane. Being the blue points the ones with largest values for this cross section, it is apparent from the figure that this is achieved at the expense of reaching the 2$\sigma$ contour in the region of lower values of $\delta \kappa_g$, in accordance to Fig.~\ref{fig-dkg-xsechh-lightscenario}. As it can be also seen from this figure, it is pointless to explore larger positive values of $\delta \kappa_g$ because they are located in a region where the corrections to the double Higgs cross section are small, as it is shown by the green points.

\begin{figure}[h!]
\centering
\includegraphics[width=0.85\textwidth]{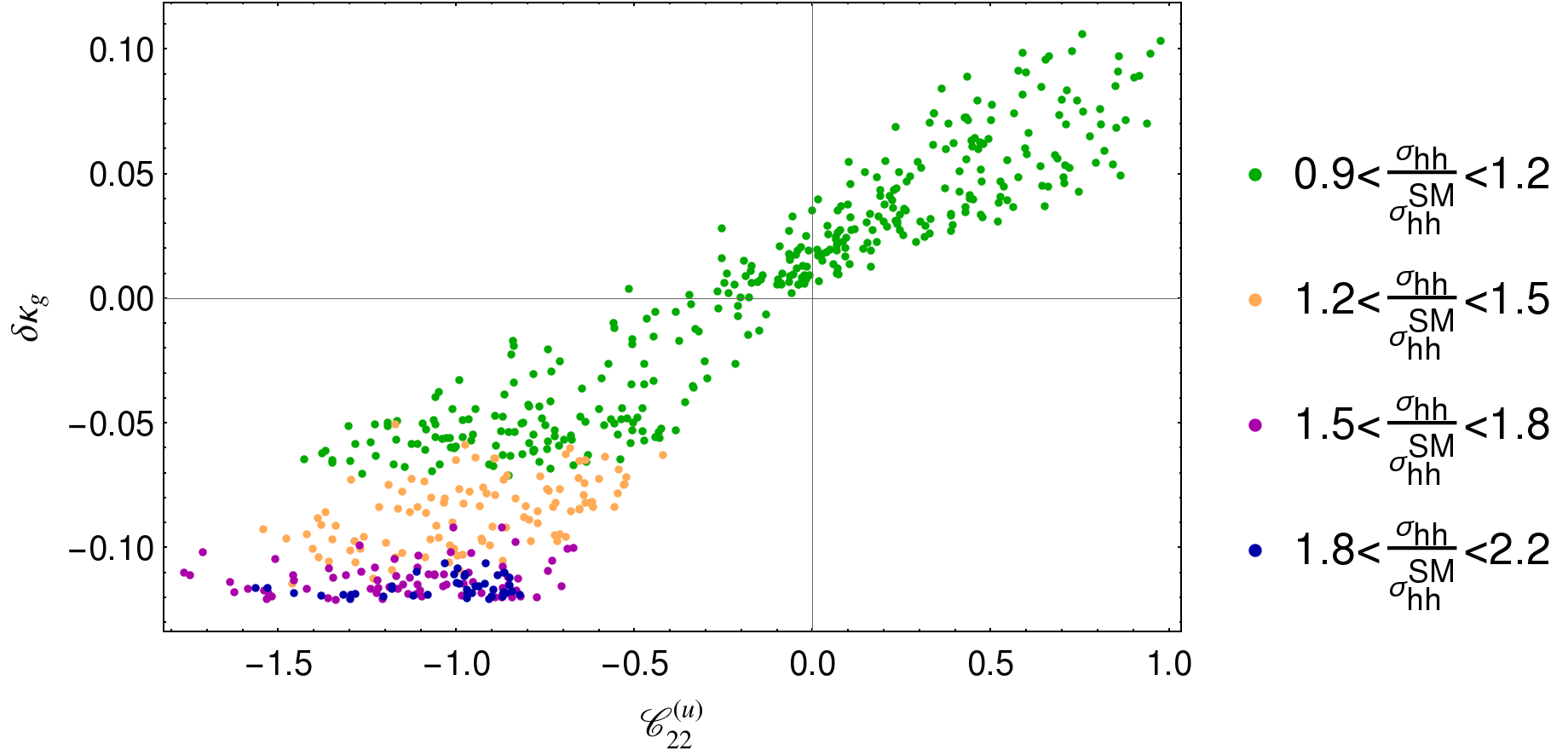}
\caption{$\delta \kappa_g$ as a function of ${\mathscr C}_{22}^{(u)}$, in colors  the values of $\sigma_{hh}/\sigma^{\rm SM}_{hh}$.}
\label{fig-c3u22-dkg-lightscenario}
\end{figure}

In Fig.~\ref{fig-c3u22-dkg-lightscenario} we plot $\delta \kappa_g$ vs. ${\mathscr C}_{22}^{(u)}$, the cubic coupling of the light up-type state. As it is shown in this figure, the blue points corresponding to the larger values of the double Higgs cross section have the larger negative values of this coupling, effect that can be followed from Eq.~(\ref{deltakappag}). In agreement with Fig.~\ref{fig-dkgamma-dkg-lightscenario}, positive values of $\delta \kappa_g$ are associated to smaller cross sections and are originated from positive values of ${\mathscr C}_{22}^{(u)}$. The horizontal break around $\delta \kappa_g \approx -0.12$ corresponds to the constraint on the single Higgs production as shown in Fig.~\ref{fig-dkgamma-dkg-lightscenario}.

\subsection{$T$ parameter}
\begin{figure}[h!]
\centering
\includegraphics[width=0.49\textwidth]{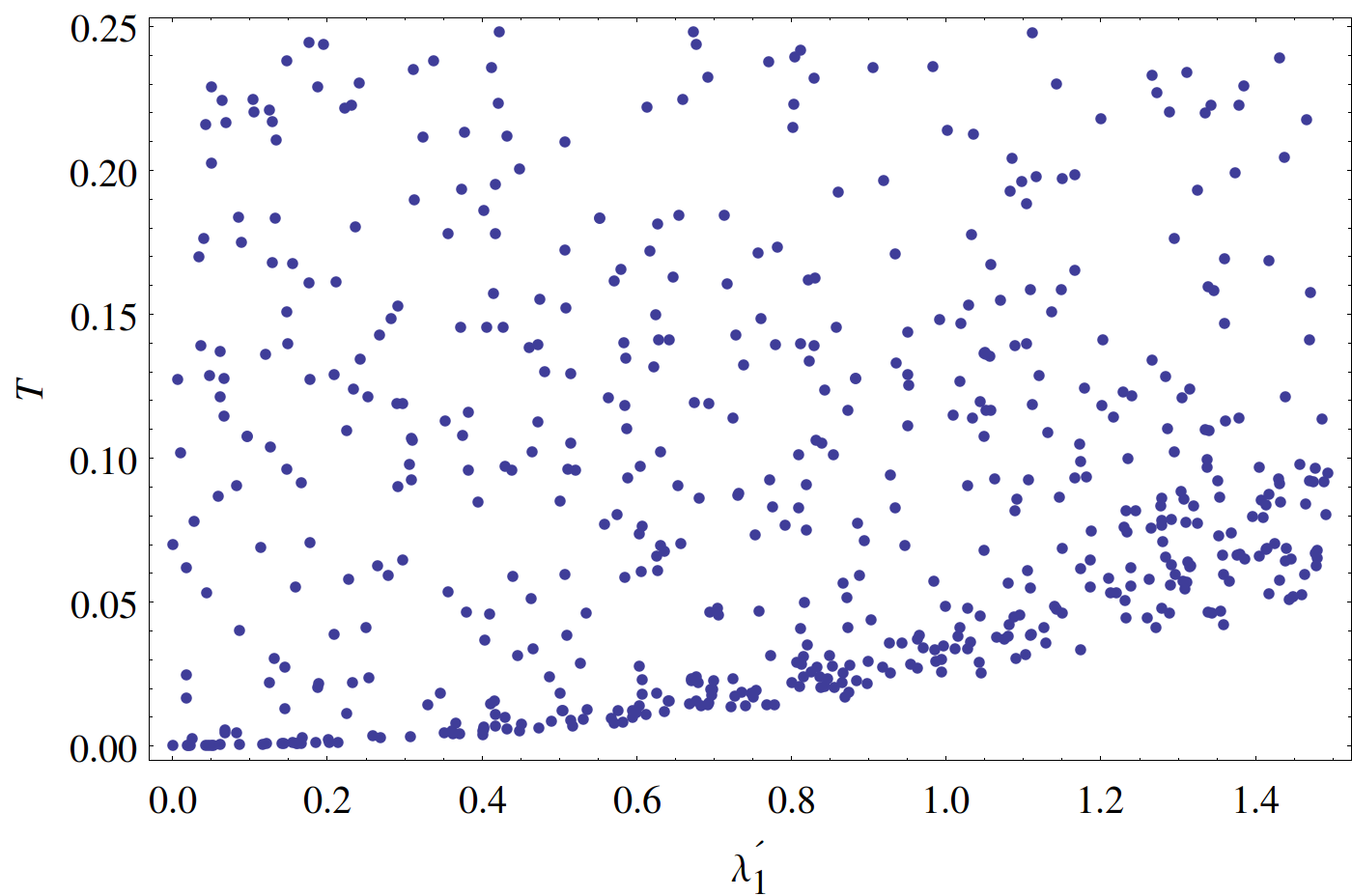}
\includegraphics[width=0.49\textwidth]{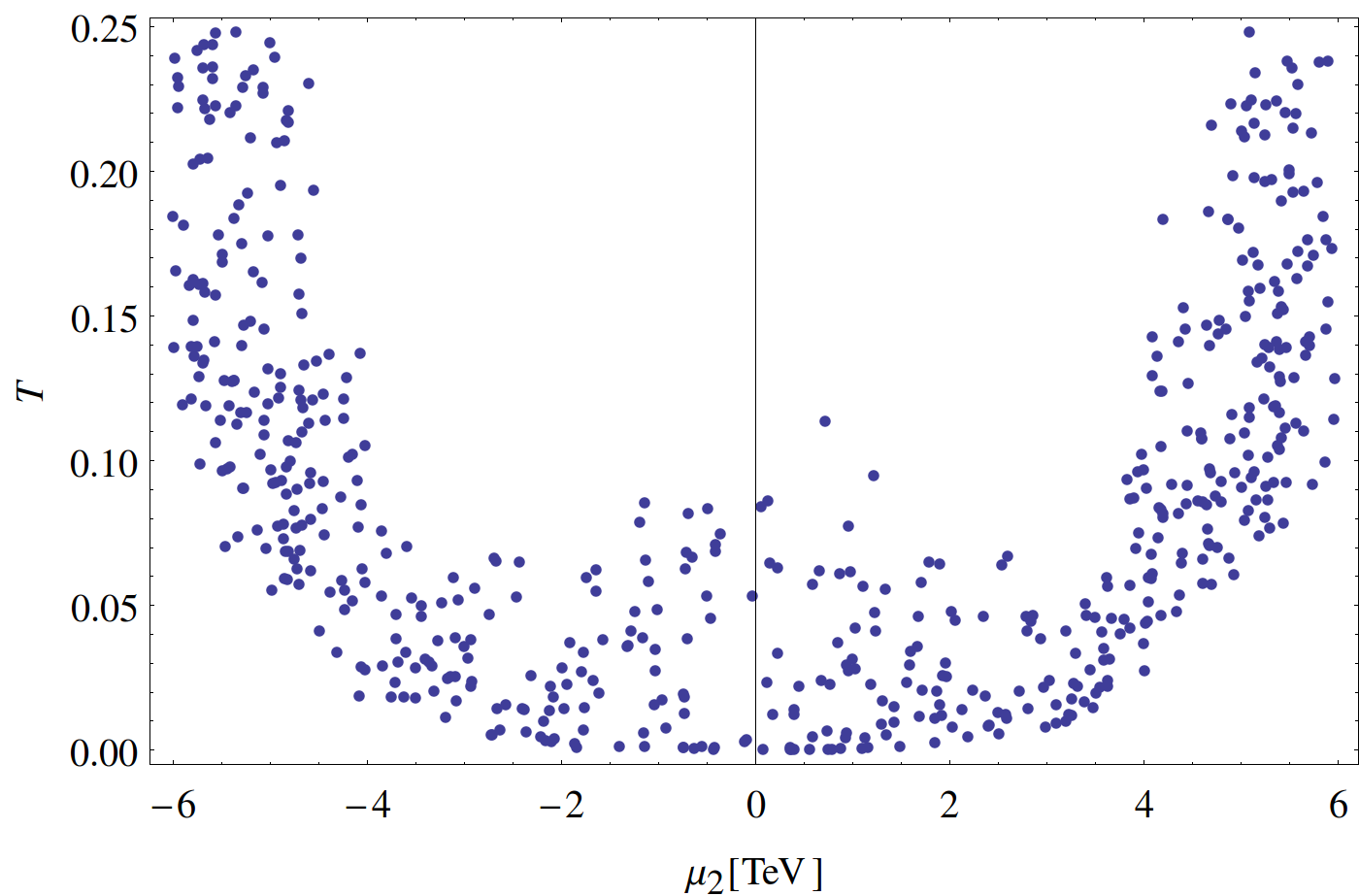}
\caption{Value of the $T$-parameter for the points of the scan. On the left dependence with $\lambda_1'$, on the right dependence with $\mu_2$.}
\label{fig-T-lightscenario}
\end{figure}

We consider now the value of the $T$-parameter for this scenario. As discussed in Sec.~\ref{sec-oblique}, $T$ is driven by $\lambda_1'$ and $\mu_i$, thus we show in Fig.~\ref{fig-T-lightscenario} this dependence. For each plot, we have checked that the approximations of Eq.~(\ref{eq-aproxT}) reproduce quite well the curve wrapping the cloud of points from below, while the points with larger values of $T$ correspond to regions of the parameter space where the approximation is not valid, due to large $\mu_2$ and/or $\lambda_1'$. For $|\mu_2|\gtrsim 6$~TeV most of the points exceed the upper limit on $T$, while one could take larger values of $\lambda_1'$, at the price of lowering $|\mu_2|$. We do not find correlation between $T$ and $\sigma_{hh}$, points with $\sigma_{hh}/\sigma_{hh}^{\rm SM}\gtrsim 1.8$ have $T\sim 0.05-0.25$ with an approximately uniform distribution.

\section{Scenario with heavy LQs}\label{sec-heavy-scenario}

We briefly discuss now the scenario 2 of Sec.~\ref{sec-BLconservation}, where up to a sign, all the LQs have the same baryon and lepton numbers. The only coupling with fermions is $y_{2\ell}$ and the simplest situation is to consider interactions with the fermions of the third generation only. Since in this case only one coupling to fermions is present, many flavor bounds, as well as bounds from direct searches, are less stringent than in the most general case. It is possible to relax even more flavor bounds by taking small couplings with the fermions of the first and second generation. 

We have made a scan on the parameters of the model with the only restriction that the couplings are perturbative, $|\mu_{1,2}|\leq 2\pi$~TeV and $\lambda_{1,2,3}, \lambda'_1\in(0,2\pi)$, and then selected points where all the LQ masses are larger than 800 GeV. The output of the scan has shown that for most of the points the corrections to the double cross section are smaller than 20\%, and for a few points they can reach 30\%. The $T$-parameter is similar to the previous scenario.

\begin{figure}[h!]
\centering
\includegraphics[width=0.9\textwidth]{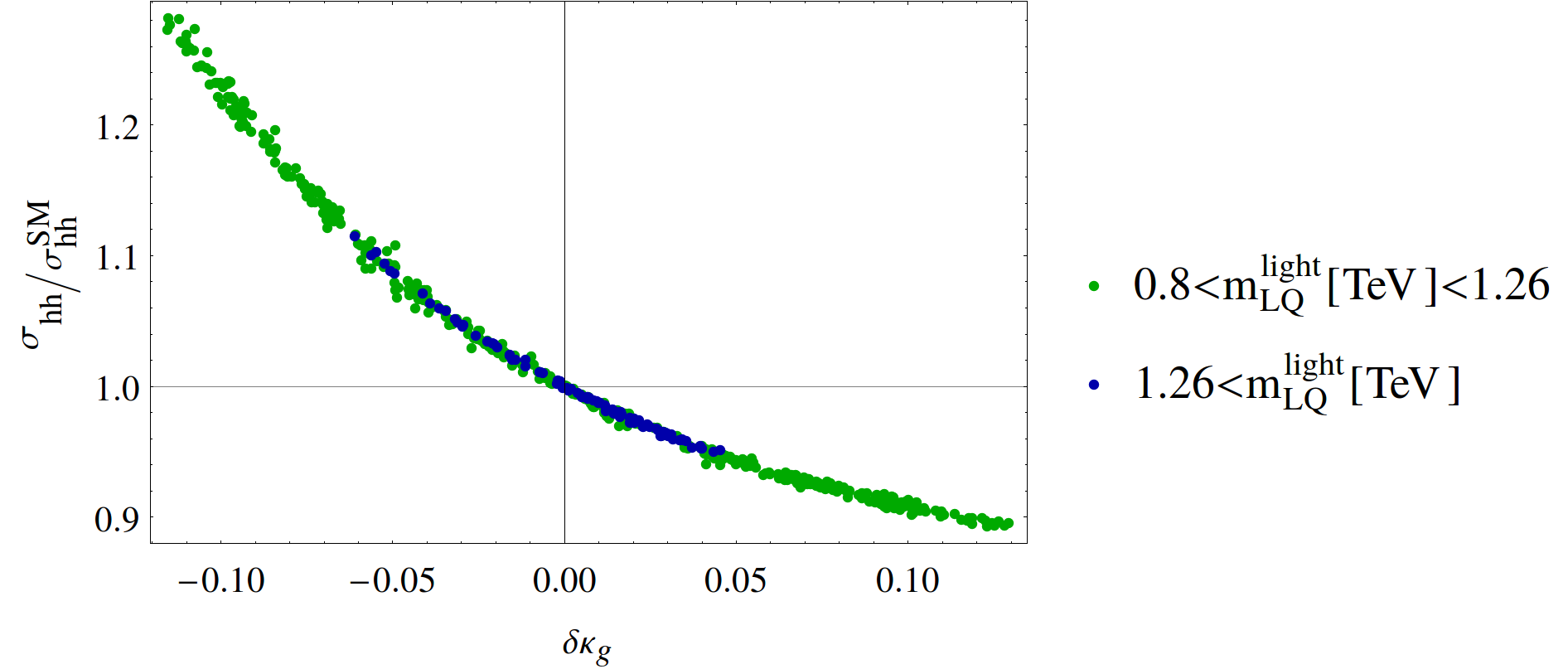}
\caption{LO $\sigma_{hh}$ normalized with respect to the SM one, as function of $\delta\kappa_g$, in color values of the lightest LQ mass $m_{\rm LQ}^{\rm light}$.}
\label{fig-dkg-xsechh-heavyscenario}
\end{figure}

In Fig.~\ref{fig-dkg-xsechh-heavyscenario} we show  $\sigma_{hh}/\sigma^{\rm SM}_{hh}$  vs. $\delta \kappa_g$. As opposed to the scenario with a light LQ, we observe a stronger correlation among both quantities which can be followed from the LET since now all the LQs are heavy. Besides, compared to that scenario, we have lower values of $\sigma_{hh}$ for similar $\delta\kappa_g$. As an example, for $\delta\kappa_g=-0.1$, the effect on the cross section is of order $20-30\%$, that is similar or smaller than the lowest values obtained in the scenario with a light LQ. However, if we compare both scenarios for similar values of $m_{\rm LQ}^{\rm light}$, we see that slightly larger values of $\sigma_{hh}$ are obtained for the heavy LQ scenario due to the fact that both $\mu_1$ and $\mu_2$ are present in this case. The blue points in this figure correspond to larger values of the lightest LQ mass, $m_{\rm LQ}^{\rm light}$. We see that for these points the correction to the SM cross section is smaller than the one corresponding to the green points for which we allow lower values of $m_{\rm LQ}^{\rm light}$, as expected.

As a consequence of the moderate correction to $\sigma^{\rm SM}_{hh}$ generated by the LQs, it is clear that this scenario will be more difficult to test in double Higgs production at the LHC.

\section{$B$ anomalies}\label{Banomalies}

Besides their interesting effects on the Higgs phenomenology, LQs can also play a role explaining a set of deviations related with flavor physics. Anomalies in the decays of $B$ mesons have been recently measured at several experiments~\cite{Lees:2013uzd,Aaij:2015yra,Hirose:2016wfn,Abdesselam:2019wac,Aaij:2021vac}, the largest ones were observed in the ratios $R_{K^{(*)}}$ and $R_{D^{(*)}}$, via neutral and charged currents, respectively. One of the most appealing possibilities to explain these deviations is the presence of LQ states at the TeV scale. There have been many proposals and studies, in particular Ref.~\cite{1808.08179} has summarized the effect of individual LQs on these observables. For the content of LQs of the present paper, only $S_1$ can accommodate $R_{K^{(*)}}$ or $R_{D^{(*)}}$, while simultaneously satisfying the bounds from other flavor observables, by adding small couplings with the second generation. For a combined explanation of both anomalies, Refs.~\cite{1608.07583,1704.05849} have shown that $S_1$ can, at best, explain them at the level of $2\sigma$.\footnote{For explanations at $1\sigma$ level see, for example, Refs.~\cite{Crivellin:2017zlb,Buttazzo:2017ixm,Crivellin:2019dwb} and references therein.} Besides, given the proper couplings, it can also accommodate the anomalous magnetic moment of the muon. 

In the scenario 1 of Sec.~\ref{sec-BLconservation} (conserving $B$ and $L$) the impact of $S_1$ on the Higgs phenomenology that we have studied is subleading. Therefore its mass and couplings with fermions can be taken to accommodate $R_{K^{(*)}}$ or $R_{D^{(*)}}$, or both of them at $2\sigma$ level. To leading order in this scenario the effects on the Higgs phenomenology and on the $B$ anomalies are decoupled, with the first ones being dominated by $\tilde R_2$ and $\bar S_1$, and the second ones by $S_1$. Whereas current limits do not exclude an explanation of the $B$ anomalies~\cite{Schmaltz:2018nls}, one may wonder to which extent this possibility might be affected by single resonant LQ production searches at the LCH in the future~\cite{Buonocore:2020erb,Greljo:2020tgv,Haisch:2020xjd}. Since the constraints arising from these searches would only impact on LQs coupled to leptons, they would not impose restrictions on the lightest LQ which in this scenario drives the increase of the double Higgs production and is mainly identified with $\bar S_1$, whose couplings to leptons we put to zero. On the other hand, this would not be the case for the heavier LQs, as $S_1$. However, since their masses can be above the TeV, one could expect that couplings to leptons of second and third generation of order 0.1-1, as those required to explain the low-energy anomalies, could satisfy the present bounds. Definitely, a dedicated analysis, that is beyond the scope of our work, is required to carefully assess an explanation of the low-energy anomalies given the bounds from single resonant production, explanation that is already in tension with the present LQ content~\cite{1608.07583,1704.05849}.\footnote{In the scenario 2 all the LQs couple to leptons and have masses around 1 TeV or above, thus the situation is similar to the heavy resonances of scenario 1.}

Generically, effects on the Higgs physics are dominated by cubic and quartic couplings between the scalar bosons, while explanations of the $B$-anomalies and the anomalous magnetic moment of the muon involve couplings to fermions. Besides, sizable effects on double Higgs production require masses below 1 TeV, whereas explanations of the anomalies can be accomplished by LQs with masses above the TeV. For these reasons, the absence of effects on double Higgs production would not discard the LQ explanation of the $B$-anomalies and/or anomalous magnetic moment of the muon.

\section{Conclusions}\label{conclusions}

In the absence of direct signals of new physics at the LHC, precision measurements in the Higgs sector could provide a portal to physics beyond the SM. In this paper we have studied the Higgs phenomenology in the presence of scalar LQs at the TeV scale. Interest in these new states has been boosted in the last years due to the anomalies observed in the decays of $B$ mesons in several experiments, as well as quite recently due to the new measurement of  the anomalous magnetic moment of the muon at Fermilab, given that LQs are one of the preferred options to explain both phenomena that seem to point to BSM physics. Besides their possible impact on flavor physics, their colored nature implies that they may also modify Higgs production at hadron colliders such as the LHC, since this production is dominated by gluon fusion, in which loops of colored particles enter. Although single Higgs production has been measured at the LHC with a precision of order 10\% in agreement with the SM, double Higgs production is naturally expected to be probed at much larger luminosities (of order 3 ab$^{-1}$) and thus there is room for ${\cal O}(1)$ modifications with respect to the SM predictions. It is however a non-trivial task for the new physics responsible for potentially large modifications in double Higgs production to produce at the same time small deviations in single Higgs production. This is one of the main issues that we have considered in this paper.

We have studied a model with three scalar LQs, a weak doublet and two singlets, with hypercharge 1/6, -2/3 and 1/3, respectively. We have shown that this set of fields, alongside their cubic and quartic interactions with the Higgs, provides an opportunity for enhancing  double Higgs production at LHC while simultaneously keeping under control  single Higgs production. We have considered the most general renormalizable potential, analyzing under some simplifying assumptions the stability of the EW vacuum and estimating the size of the couplings where the perturbative expansions are under control. At the leading order in which we work, the contributions from the LQs to Higgs production are decoupled from the LQs interactions with SM fermions. We have determined the conditions for baryon and lepton number conservation from the LQs interactions with fermions  and we have estimated the bounds on these interactions from flavor physics. Another important constraint that we have considered comes from the $T$-parameter, that receives contributions from the cubic and some of the quartic interactions that induce a splitting between up- and down-type LQs. These contributions have been computed and we have selected the regions of the parameter space  that satisfy the bounds. It is worth mentioning that when considering the bounds from the current LHC measurements on single Higgs production and decay, we added the LQs contributions to Higgs decays that involve loop-level generated couplings to the weak sector such as to diphotons or $Z\gamma$, finding, in agreement with the literature, that these corrections are in general below the current bounds and are less constrained than the gluon fusion contributions to single Higgs production.

A full numerical calculation of the cross section of double Higgs production at LHC, at leading order, is performed in the model. Comparing our results with similar studies in the literature, both in presence of LQs and in the presence of supersymmetric particles,  we find good agreement in the corresponding parameter space. Since the full calculation must be carried at the numerical level, is very expensive computational-wise and sometimes even the physics may be hard to extract, we made extensive use of the Higgs low energy theorems for heavy LQ masses. Considering a correction to the interference term between the new physics and the SM by a factor $\simeq 2$ due to the relatively low top quark mass compared to others scales of the process, the analysis using the LET allowed us to understand several interesting relations, as the dependence of the cross sections with some of the parameters of the potential and with some physical masses and couplings. It also helped us in the search for the desired regions of the parameter space with better efficiency.  In this context, we were able to find a rather simple relation between $\sigma_{hh}$ and $\delta \kappa_g$ that is valid in the limit of dominance of cubic couplings and describes well the trend shown in the full calculation with deviations  arising mostly from  quartic couplings but with a clear discrepancy in the regions of small LQ masses in which the LET breaks down. 

Our study focuses on two different scenarios. The first one consists of a light LQ with a mass of order $400-600$ GeV and fermionic interactions such that it only decays to dijets, evading the strongest bounds from direct detection, while the other LQs considered have masses larger than 800 GeV satisfying all direct search bounds. Following the indications from LET and using the full calculation, we find regions of the parameter space where the cross section can be more than two times the SM one, satisfying at the same time all constraints, in particular the ones coming from Higgs single production. In these regions of large enhancement, the cross section is dominated by the contribution from the light state. Using the expected sensitivities for double Higgs production from both ATLAS and CMS at the high luminosity LHC, we estimate that hints of the largest enhancements to di-Higgs production in this scenario should start to become apparent at luminosities $\mathcal{L}\sim 2$ ab$^{-1}$.  The second scenario considers all the LQs to have masses larger than 800 GeV. In this case the corrections to the double Higgs production cross section, compatibles with the bounds from single Higgs production, are in general not larger than 20\% than that of the SM, although in some cases they may reach 30\%. 

Finally, we notice that the separation between the LQs interactions with the Higgs sector (and their consequences on Higgs production) and the LQs interactions with the fermionic sector (which would provide higher order corrections to Higgs productions and are expected to be small) implies that an explanation of the $B$ anomalies, and/or the anomalous magnetic moment of the muon, together with an enhancement of double Higgs productions, can be accomplished in the model we study. The best scenario for the explanation of the $B$ anomalies with scalar LQs contains a weak triplet and a singlet with hypercharges 1/3. Likewise, there is a possibility to address $(g-2)_{\mu}$ within a similar scenario. Although the lack of any relevant impact on the double Higgs production would not prevent to explain the $B$-anomalies and/or $(g-2)_{\mu}$ in terms of LQs, it may be interesting to perform a complete study for the possible explanations of these anomalies within our setup and we leave it for future work.

\acknowledgments

We thank Javier Mazzitelli for assistance on the numerical calculations as well as Carlos Wagner for useful discussions. L.D. thanks IFLP for its hospitality during the accomplishment of this work. This work has been partially supported by CONICET and ANPCyT projects PIP 11220150100299CO, PICT 2016-0164, PICT 2017-0802, PICT 2017-2751 and PICT 2018-03682. M.E. acknowledges the funding from the Academy of Finland, project 308301.

\appendix

\section{Higgs-LQs interactions in the physical basis}
\label{sec:appendix:Higgs-leptoquarksphysicalbasis}

After the EWSB, the potentials $V_3$ and $V_4$ provide the Higgs-LQ interaction. These could be written as next

\begin{align*}
    V_3+V_4 \rightarrow
    h {\footnotesize \begin{pmatrix}
    \tilde{R}^d_2 \\
    S_1^*
    \end{pmatrix}}^\dagger
    & C^{(d)}
    {\footnotesize \begin{pmatrix}
    \tilde{R}^d_2 \\
    S_1^*
    \end{pmatrix}} +
    h {\footnotesize \begin{pmatrix}
    \tilde{R}^u_2 \\
    \bar{S}_1^*
    \end{pmatrix}}^\dagger
    C^{(u)}
    {\footnotesize \begin{pmatrix}
    \tilde{R}^u_2 \\
    \bar{S}_1^*
    \end{pmatrix}}
     + h^2 {\footnotesize \begin{pmatrix}
    \tilde{R}^d_2 \\
    S_1^*
    \end{pmatrix}}^\dagger
    Q^{(d)}
    {\footnotesize \begin{pmatrix}
    \tilde{R}^d_2 \\
    S_1^*
    \end{pmatrix}} +
    h^2 {\footnotesize \begin{pmatrix}
    \tilde{R}^u_2 \\
    \bar{S}_1^*
    \end{pmatrix}}^\dagger
    Q^{(u)}
    {\footnotesize \begin{pmatrix}
    \tilde{R}^u_2 \\
    \bar{S}_1^*
    \end{pmatrix}}
\end{align*}
where we use the $\phi = {(v + h)}/{\sqrt{2}}$ prescription and

\begin{align*}
\begin{array}{cc}
        C^{(u)} = \begin{pmatrix}
        v (\lambda_1+\lambda_1') & \dfrac{\mu_2}{\sqrt{2}} \\
        \dfrac{\mu_2}{\sqrt{2}}  & v \lambda_3
        \end{pmatrix} & \; \;
        C^{(d)} = \begin{pmatrix}
        v (\lambda_1-\lambda_1') & \dfrac{\mu_1}{\sqrt{2}} \\
        \dfrac{\mu_1}{\sqrt{2}}   & v \lambda_2
        \end{pmatrix} \\[30pt]
\end{array}
\end{align*}
\begin{align*}
\begin{array}{cc}
    Q^{(u)} = \begin{pmatrix}
        \dfrac{\lambda_1+\lambda_1'}{2} & 0 \\
        0 & \dfrac{\lambda_3}{2}
        \end{pmatrix} & \; \; \; \; \; \;
        Q^{(d)} = \begin{pmatrix}
        \dfrac{\lambda_1-\lambda_1'}{2} & 0 \\
        0 & \dfrac{\lambda_2}{2}
        \end{pmatrix}
\end{array}
\end{align*}

Once mass matrices are simultaneously diagonalized, the EWSB picture of the combination $V_3$ and $V_4$ can be rewritten in terms of the physical fields, $\chi^u_1$, $\chi^u_2$, $\chi^d_1$ and $\chi^d_2$ as follows

\begin{align*}
    V_3+V_4 \rightarrow
    h {\footnotesize \begin{pmatrix}
    \chi^d_1 \\
    \chi^d_2
    \end{pmatrix}}^\dagger
    & \mathscr{C}^{(d)}
    {\footnotesize \begin{pmatrix}
    \chi^d_1 \\
    \chi^d_2
    \end{pmatrix}} +
    h {\footnotesize \begin{pmatrix}
    \chi^u_1 \\
    \chi^u_2
    \end{pmatrix}}^\dagger
    \mathscr{C}^{(u)}
    {\footnotesize \begin{pmatrix}
    \chi^u_1 \\
    \chi^u_2
    \end{pmatrix}}     
    + h^2 {\footnotesize \begin{pmatrix}
    \chi^d_1 \\
    \chi^d_2
    \end{pmatrix}}^\dagger
    \mathcal{Q}^{(d)}
    {\footnotesize \begin{pmatrix}
    \chi^d_1 \\
    \chi^d_2
    \end{pmatrix}} +
    h^2 {\footnotesize \begin{pmatrix}
    \chi^u_1 \\
    \chi^u_2
    \end{pmatrix}}^\dagger
    \mathcal{Q}^{(u)}
    {\footnotesize \begin{pmatrix}
    \chi^u_1 \\
    \chi^u_2
    \end{pmatrix}}
\end{align*}
The new matrices are defined from $C^{(u)}$, $C^{(d)}$, $Q^{(u)}$ and $Q^{(d)}$, and the rotation matrices presented in Eqs.~(\ref{eq-rot1}) and~(\ref{eq-rot2})

\begin{equation*}
    \mathscr{C}^{(u)} = U_u^{-1} C^{(u)} U_u
\end{equation*}
\begin{equation*}
    \mathscr{C}^{(d)} = U_d^{-1} C^{(d)} U_d
\end{equation*}
\begin{equation*}
    \mathcal{Q}^{(u)} = U_u^{-1} Q^{(u)} U_u
\end{equation*}
\begin{equation*}
    \mathcal{Q}^{(d)} = U_d^{-1} Q^{(d)} U_d
\end{equation*}
The matrix elements of these matrices are the physical field coupling constants that rule the interaction between the LQ mass eigenstate fields and the Higgs boson. They could be explicitly expressed in terms of the gauge parameters and the mixing angles as follows

\begin{align*}
    \mathscr{C}^{(u)}_{11} &= v \Big[\lambda_3 + (\lambda_1+\lambda_1'-\lambda_3 ) \sin ^2(\theta_u )\Big] -\frac{\mu _2}{\sqrt{2}} \sin(2 \theta_u )
    \\
    \mathscr{C}^{(u)}_{12} &= \mathscr{C}^{(u)}_{21} = \frac{\mu _2}{\sqrt{2}}\cos (2 \theta _u)-\frac{1}{2} v (\lambda _1'+\lambda _1-\lambda _3) \sin
   (2 \theta _u)
   \\
   \mathscr{C}^{(u)}_{22} &= v \Big[\lambda_3 + (\lambda_1+\lambda_1'-\lambda_3 ) \cos ^2(\theta_u )\Big] +\frac{\mu _2}{\sqrt{2}} \sin(2 \theta_u ) \\
\\
    \mathscr{C}^{(d)}_{11} &= v \Big[\lambda_2 + (\lambda_1-\lambda_1'-\lambda_2 ) \sin ^2(\theta_d)\Big] -\frac{\mu_1}{\sqrt{2}} \sin(2 \theta_d )
    \\
    \mathscr{C}^{(d)}_{12} &= \mathscr{C}^{(d)}_{21} = \frac{\mu _1}{\sqrt{2}} \cos (2 \theta _d)-\frac{1}{2} v (\lambda _1'-\lambda _1-\lambda _2) \sin(2 \theta _d)
   \\
   \mathscr{C}^{(d)}_{22} &= v \Big[\lambda_2 + (\lambda_1-\lambda_1'-\lambda_2 ) \cos ^2(\theta_d )\Big] +\frac{\mu _1}{\sqrt{2}} \sin(2 \theta_d ) \\
\\
  \mathcal{Q}^{(u)}_{11} &= \frac{1}{2} (\lambda _1'+\lambda _1-\lambda_3) \sin ^2(\theta_u)+ \frac{\lambda_3}{2}
    \\
    \mathcal{Q}^{(u)}_{12} &= \mathcal{Q}^{(u)}_{21} = - \frac{1}{4} (\lambda _1'+\lambda _1-\lambda _3) \sin (2 \theta_u)
   \\
   \mathcal{Q}^{(u)}_{22} &= \frac{1}{2} (\lambda _1'+\lambda _1-\lambda_3) \cos ^2(\theta_u)+ \frac{\lambda_3}{2} \\
\\
    \mathcal{Q}^{(d)}_{11} &= \frac{1}{2} (\lambda _1'-\lambda _1-\lambda_2) \sin ^2(\theta_d)+ \frac{\lambda_2}{2}
    \\
    \mathcal{Q}^{(d)}_{12} &= \mathcal{Q}^{(d)}_{21} = - \frac{1}{4} (\lambda _1'-\lambda _1-\lambda _2) \sin (2 \theta_d)
   \\
   \mathcal{Q}^{(d)}_{22} &= \frac{1}{2} (\lambda _1'-\lambda _1-\lambda_2) \cos ^2(\theta_d)+ \frac{\lambda_2}{2}
\end{align*}
It is worth noting that, since we assume all parameters are real, $\mathscr{C}^{(u)}$, $\mathscr{C}^{(d)}$, $\mathcal{Q}^{(u)}$ and $\mathcal{Q}^{(d)}$ turn out to be symmetrical.

\section{Perturbative expansion}
\label{perturbativeexpansion}

In this Appendix we make some estimations of the radiative corrections and bounds on the couplings from perturbativity.

Neglecting the mixing between the LQs, the cubic couplings contribute to double Higgs production but not to single Higgs production at LHC, therefore it is interesting to consider large cubic couplings. This couplings can induce large radiative corrections, and eventually break the perturbative expansion. Thus we study NDA estimates of the allowed size of the cubic couplings. 

\begin{figure}[h!]
\centering
\includegraphics[width=0.25\textwidth]{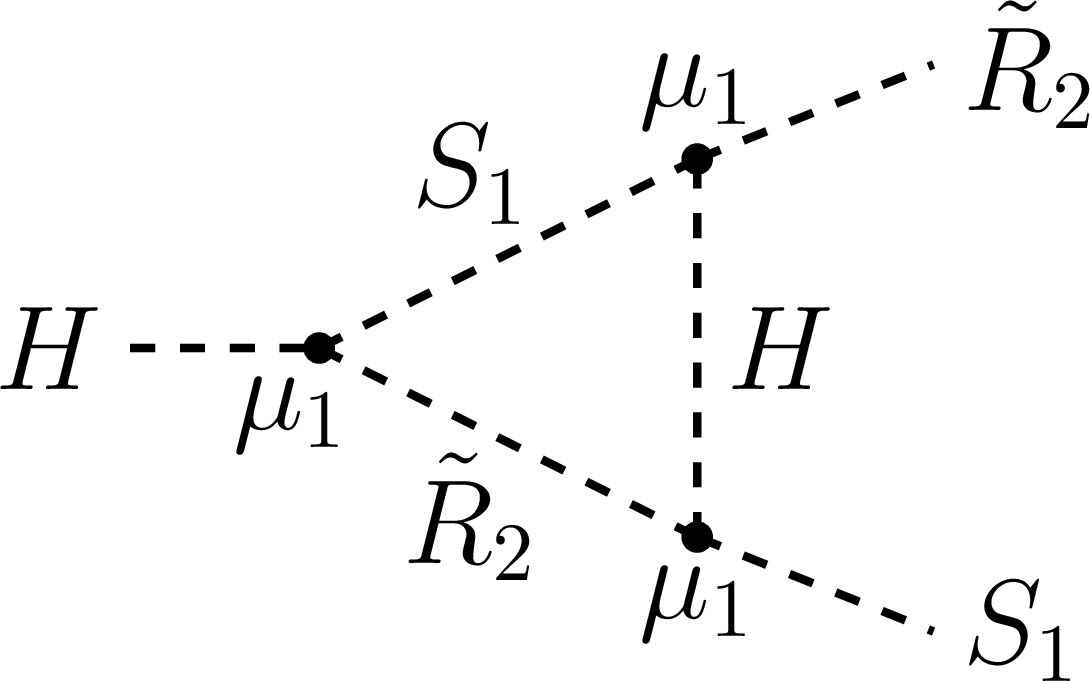}\hspace*{1.35cm}
\includegraphics[width=0.25\textwidth]{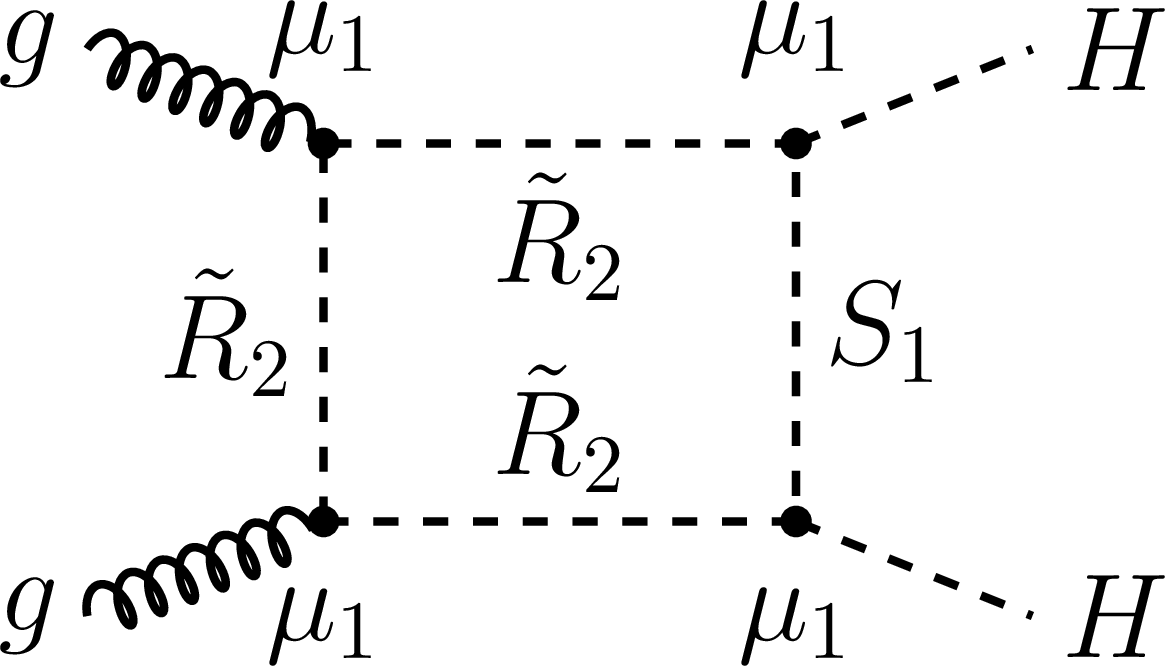}\hspace*{1.35cm}
\includegraphics[width=0.33\textwidth]{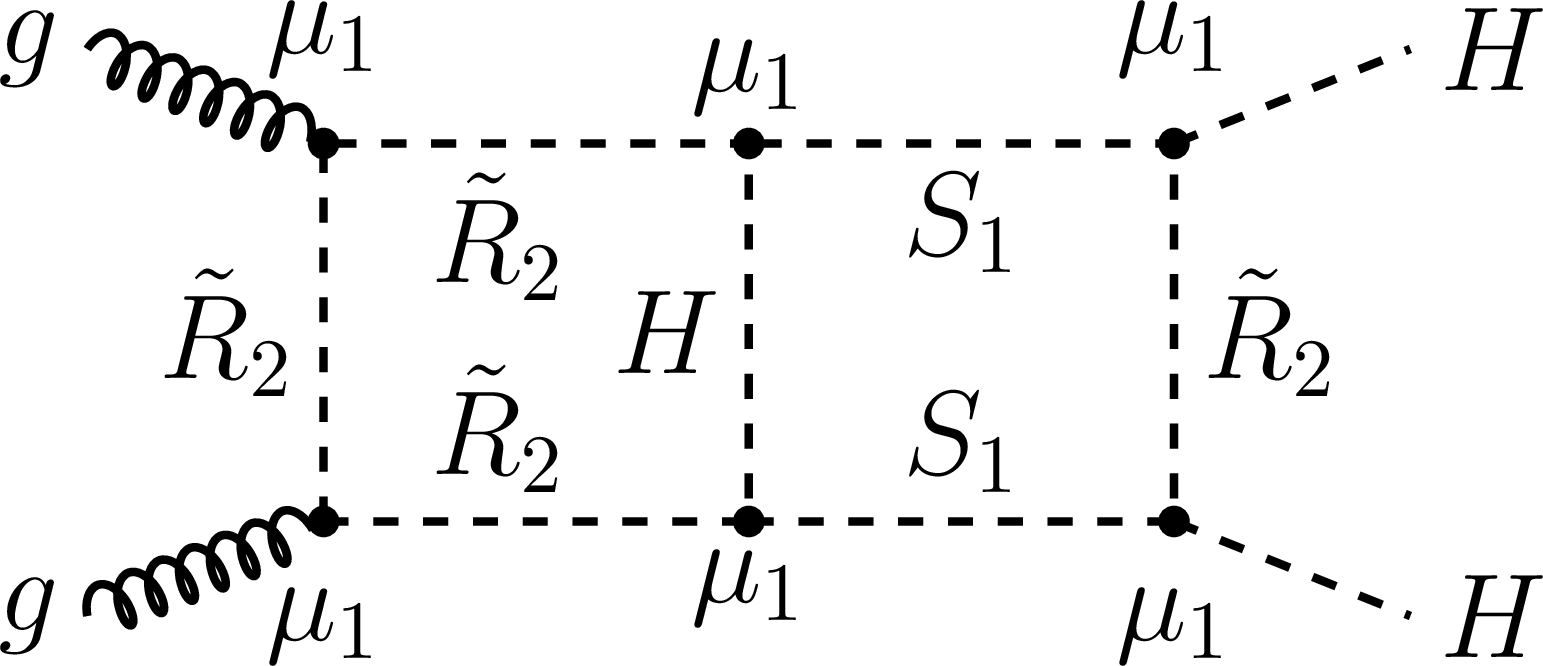}
\caption{One Feynman diagram correcting the cubic coupling at one loop (left), and contributions at one and two loops to $gg\to HH$.}
\label{fig-NDA}
\end{figure}

We consider the terms of $V_3$ in Eq.~(\ref{eq-V3}), that are corrected at one loop by the Feynman diagrams of Fig.~\ref{fig-NDA}(left).

At this level the correction to $\mu$-itself is given by
\begin{align*}
&\mu_1^3\int\frac{d^4p}{(2\pi)^4}\frac{1}{p_H^2-m_H^2}\frac{1}{p_{S_1}^2-m_{S_1}^2}\frac{1}{p_{\tilde R_2}^2-m_{\tilde R_2}^2}=\nonumber
\\
&\frac{-i\mu_1^3}{16\pi^2}\frac{m_H^2m_{\tilde R_2}^2\log\frac{m_H}{m_{\tilde R_2}}+m_{S_1}^2m_{\tilde R_2}^2\log\frac{m_{\tilde R_2}}{m_{S_1}}+m_H^2m_{S_1}^2\log\frac{m_{S_1}}{m_H}}{(m_H^2-m_{\tilde R_2}^2)(m_H^2-m_{S_1}^2)(m_{\tilde R_2}^2-m_{S_1}^2)} \, ,
\end{align*}
where we have considered the limit of vanishing external momenta. Taking the limit of $m_{\rm LQ}\gg m_H$ and demanding the one loop correction to be smaller than the tree level coupling we obtain
\begin{equation}
\mu_1\lesssim 4\pi\left(\frac{m_{\tilde R_2}^2-m_{S_1}^2}{\log m_{\tilde R_2}/m_{S_1}}\right)^{1/2} \ . \nonumber
\label{eq-boundNDA}
\end{equation}
For degenerate masses it reduces to $\mu_1\lesssim 4\pi m_{\rm LQ}$, whereas in the case of a large splitting $m_{\rm heavy}\gg m_{\rm light}$: $\mu_1\lesssim 4\pi m_{\rm heavy}/(\log m_{\rm heavy}/m_{\rm light})^{1/2}$. A similar result is obtained for $\mu_2$, changing $S_1$ by $\bar S_1$.

Large cubic couplings can also have an impact on the cross section of double Higgs production. In Fig.~\ref{fig-NDA} we show diagrams at one and two loops with the largest number of cubic interactions. These integrals can be computed analytically in the limit of negligible Higgs mass and zero external momentum, leading to: $I^{(2)}/I^{(1)}=(1/2)(\mu_1/4\pi m_{\tilde R_2})^2$, where the superindex labels the number of loops. We have also computed these integrals numerically for external momentum of order 500~GeV and different angles. For the regions of the parameter considered in the phenomenological analysis we find this ratio to be in general $\lesssim 1\%$, with a few points being of order 5-10\%. A full two loop calculation of these corrections is beyond the scope of this work. We take cubic couplings below 1/3 of the bound given above.

\section{Form factors}
\label{sec:appendix:formfactos}

The leading order form factors that parametrize the cross section of the reaction

\begin{equation*}
    g(p_1) + g(p_2) \rightarrow h(p_3) + h(p_4)    
\end{equation*}
are presented in terms of the Mandelstam variables

\begin{align*}
    \hat{s} &= (p_1 + p_2)^2 = (p_3 + p_4)^2
    \\
    \hat{t} &= (p_1 - p_3)^2 = (p_4 - p_2)^2
    \\
    \hat{u} &= (p_1 - p_4)^2 = (p_3 - p_2)^2,
\end{align*}
the masses of the quark in the loop $m_q$, in the case of the Standard Model contributions, and $m_i$ of the mass-eigenstate LQs, and the effective couplings $\mathscr{C}^{(k)}_{ij}$ and $\mathcal{Q}^{(k)}_{ij}$.

\begin{align*}
    \mathcal{F}_\triangle &= \frac{12 m_h^2 m_q^2 }{(\hat{s}-m_h^2)} \Big[2+\big(4 m_q^2-\hat{s}\big)
    \text{C}^{00}_{qqq}(\hat{s})\Big] \\
\\
    \mathcal{G}_\triangle &= 0 \\
\\
    \mathcal{F}_\square &= - \frac{2  m_q^2}{\hat{s}} \bigg[ -4\hat{s} -8  m_q^2~ \hat{s}~ \text{C}^{00}_{qqq}(\hat{s}) 
    \\
    & -  2\big(4  m_q^2-m_h^2\big) \Big\{2 \big(m_h^2-\hat{t}\big) \text{C}^{h0}_{qqq}(\hat{t}) +2 \big(m_h^2-\hat{u}\big) \text{C}^{h0}_{qqq}(\hat{u}) 
    \\
    & - \big(m_h^4-\hat{t}\hat{u}\big) \text{D}^{h0h0}_{qqqq}(\hat{t},\hat{u}) \Big\}  +2  m_q^2 \hat{s} \big(2 m_h^2-8  m_q^2+\hat{s}\big) 
    \\
    & \times \Big\{ \text{D}^{hh00}_{qqqq}(\hat{s},\hat{t}) + \text{D}^{hh00}_{qqqq}(\hat{s},\hat{u}) + \text{D}^{h0h0}_{qqqq}(\hat{t},\hat{u}) \Big\}\bigg] \\
\\
    \mathcal{G}_\square &= \frac{2  m_q^2}{m_h^4 - \hat{t} \hat{u}} \bigg[ (2m_h^4 - \hat{t}^2 - \hat{u}^2) (8  m_q^2 - \hat{t} - \hat{u}) \text{C}^{hh}_{qqq}(\hat{s}) 
    \\
    & +  (m_h^4 - 8 m_q^2 \hat{t}\, + \hat{t}^2\,) \Big\{2 (m_h^2-\hat{t}\,) \text{C}^{0h}_{qqq}(\hat{t})\,\, - \hat{s} \text{C}^{00}_{qqq}(\hat{s}) + \hat{s} t\, \text{D}^{00hh}_{qqqq}(\hat{s},\hat{t})\Big\} 
    \\
    & +  (m_h^4 - 8  m_q^2 \hat{u} + \hat{u}^2) \Big\{2 (m_h^2-\hat{u}) \text{C}^{0h}_{qqq}(\hat{u}) - \hat{s} \text{C}^{00}_{qqq}(\hat{s}) + \hat{s} \hat{u} \text{D}^{00hh}_{qqqq}(\hat{s},\hat{u})\Big\}
    \\
    & +  2 m_q^2~ (m_h^4-\hat{t} \hat{u})~ (8  m_q^2 - \hat{t} - \hat{u})
    \\
    & \times \Big\{\text{D}^{0h0h}_{qqqq}(\hat{t},\hat{u}) + \text{D}^{00hh}_{qqqq}(\hat{s},\hat{t}) + \text{D}^{00hh}_{qqqq}(\hat{s},\hat{u})\Big\} \bigg] \\
\\
    \mathcal{F}^{(1)}_{\chi^{k}_i} &= -\frac{6 {\mathscr{C}^{(k)}_{ii}} m_h^2 v }{\hat{s}-m_h^2} \Big[1 + 2m_i^2 \text{C}^{00}_{iii}(\hat{s})\Big] \\
\\   
    \mathcal{G}^{(1)}_{\chi^k_i} &= 0 \\
\\
    \mathcal{F}^{(2)}_{\chi^{k}_i} &= -4 { \mathcal{Q}^{(k)}_{ii}} v^2 \Big[1 +2m_i^2 \text{C}^{00}_{iii}(\hat{s})\Big] \\
\\
    \mathcal{G}^{(2)}_{\chi^k_i} &= 0 \\
\\
    \mathcal{F}^{(3)}_{\chi^k_i \chi^k_j} &= 6 {\mathscr{C}^{(k)}_{ij}}^2 v^2 \bigg[\text{C}^{hh}_{iji}(\hat{s}) + \text{C}^{hh}_{jij}(\hat{s})\bigg] \\
\\
    \mathcal{G}^{(3)}_{\chi^k_i \chi^k_j} &= - \frac{2 {\mathscr{C}^{(k)}_{ij}}^2 v^2}{m_h^4 - \hat{t} \hat{u}} 
    (2 \hat{t} \hat{u} - 2 m_h^4 + m_h^2 \hat{s})
    \Big\{\text{C}^{hh}_{iji}(\hat{s}) + \text{C}^{hh}_{jij}(\hat{s})\Big\} \\
\\   
    \mathcal{F}^{(4)}_{\chi^k_i \chi^k_j} &= -\frac{2 {\mathscr{C}^{(k)}_{ij}}^2 v^2}{\hat{s}} \bigg[2(m_h^2-\hat{t}) \Big\{\text{C}^{h0}_{ijj}(\hat{t})+\text{C}^{h0}_{jii}(\hat{t})\Big\} + 2(m_h^2-\hat{u}) \Big\{\text{C}^{h0}_{ijj}(\hat{u})+\text{C}^{h0}_{jii}(\hat{u})\Big\}
    \\
    & + 2 m_i^2 \hat{s} \Big\{\text{D}^{hh00}_{ijii}(\hat{s},\hat{t}) + \text{D}^{hh00}_{ijii}(\hat{s},\hat{u})\Big\} + 2 m_j^2 \hat{s} \Big\{\text{D}^{hh00}_{jijj}(\hat{s},\hat{t}) + \text{D}^{hh00}_{jijj}(\hat{s},\hat{u})\Big\}
    \\
    & + \big((m_i^2 + m_j^2) \hat{s} + \hat{t} \hat{u} - m_h^4\big) \Big\{\text{D}^{h0h0}_{ijji}(\hat{u},\hat{t}) + \text{D}^{h0h0}_{jiij}(\hat{u},\hat{t})\Big\}
    \bigg] 
    \\
    & - 6 {\mathscr{C}^{(k)}_{ij}}^2 v^2 \bigg[\text{C}^{hh}_{iji}(\hat{s}) + \text{C}^{hh}_{jij}(\hat{s})\bigg] \\
\\
    \mathcal{G}^{(4)}_{\chi^k_i \chi^k_j} &= - \frac{2 { \mathscr{C}^{(k)}_{ij}}^2 v^2}{m_h^4 - \hat{t} \hat{u}} \bigg[ (2 m_h^4 - \hat{t}^2 - \hat{u}^2) \Big\{ \text{C}^{hh}_{iji}(\hat{s}) + \text{C}^{hh}_{jij}(\hat{s}) \Big\}
    \\
    & + \big((m_i^2 - m_j^2) (m_h^4 - \hat{t} \hat{u} - 2 \hat{s} \hat{u}) - \hat{s} \hat{u}^2\big) \Big\{ \text{D}^{hh00}_{ijii}(\hat{s},\hat{u}) + \text{D}^{hh00}_{jijj}(\hat{s},\hat{u}) \Big\} 
    \\
    & + \big((m_i^2 - m_j^2) (m_h^4 - \hat{t} \hat{u} - 2 \hat{s}\, \hat{t}) - \hat{s}\, \hat{t}^2\big) \Big\{ \text{D}^{hh00}_{ijii}(\hat{s},\hat{t}) + \text{D}^{hh00}_{jijj}(\hat{s},\hat{t}) \Big\} 
    \\
    & + \hat{s} \big((t + \hat{u}) + 2 (m_i^2 - m_j^2)\big) \text{C}^{00}_{iii}(\hat{s}) + \hat{s} \big((\hat{t} + \hat{u}) + 2 (m_j^2 - m_i^2)\big) \text{C}^{00}_{jjj}(\hat{s}) 
    \\
    & - 2 (m_h^2 - \hat{t})~ \hat{t}~  \Big\{\text{C}^{h0}_{ijj}(\hat{t}) + \text{C}^{h0}_{jii}(\hat{t})\Big\} - 2 (m_h^2 - \hat{u})~ \hat{u}~ \Big\{\text{C}^{h0}_{ijj}(\hat{u}) + \text{C}^{h0}_{jii}(\hat{u})\Big\}
    \\
    & + \big((m_i^2 + m_j^2) (m_h^4 - \hat{t} \hat{u}) - \hat{s} (m_i^2 - m_j^2)^2\big) \times 
    \\
    & \hspace{2cm} \times \Big\{ \text{D}^{h0h0}_{ijji}(\hat{t},\hat{u}) + \text{D}^{hh00}_{ijii}(\hat{s},\hat{t}) + \text{D}^{hh00}_{ijii}(\hat{s},\hat{u}) +
    \\
    &  \hspace{4cm} + \text{D}^{h0h0}_{jiij}(\hat{t},\hat{u}) + \text{D}^{hh00}_{jijj}(\hat{s},\hat{t}) + \text{D}^{hh00}_{jijj}(\hat{s},\hat{u}) \Big\} \bigg]
    \\
    & + \frac{2 { \mathscr{C}^{(k)}_{ij}}^2 v^2}{m_h^4 - \hat{t} \hat{u}} (2 \hat{t}\hat{u} -2 m_h^4 + m_h^2 \hat{s}) \bigg[ \text{C}^{hh}_{iji}(\hat{s}) + \text{C}^{hh}_{jij}(\hat{s}) \bigg] 
\end{align*}
Here, the amounts $\text{C}^{ab}_{ijk}(\hat{z})$ and $\text{D}^{abcd}_{ijkl}(\hat{w},\hat{z})$ are a short notation to denote the scalar Passarino-Veltman integrals C$_0$ and D$_0$

\begin{align*}
    \text{C}^{ab}_{ijk}(\hat{z})&=\text{C}_0(m_a^2,m_b^2,\hat{z},m_i^2,m_j^2,m_k^2) \\
\\
    \text{D}^{abcd}_{ijkl}(\hat{w},\hat{z})&=\text{D}_0(m_a^2,m_b^2,m_c^2,m_d^2,\hat{w},\hat{z},m_i^2,m_j^2,m_k^2,m_l^2)
\end{align*}


\bibliographystyle{unsrt}

\end{document}